%% file: zastatHEP.tex
\documentclass[11pt]{article}
\usepackage{sint,macros,macros_static,cite}
\usepackage{amssymb}
\usepackage{epsfig}
\begin{document}
\input title.tex
\input sect1.tex

\input sect2.tex
\input sect3.tex

\input sect4.tex
\input sect5.tex

\input sect6.tex
\begin{appendix}
\input apdxa.tex
\input apdxb.tex

\input apdxc.tex

\end{appendix}
\bibliography{lattice_ALPHA}
\bibliographystyle{h-elsevier3}
\end{document}

%% file: title.tex
\begin{titlepage}

\begin{flushright}
MS-TP-03-1\\
SHEP 03/01\\
DESY 03-007\\
SFB/CPP-03-01
\end{flushright}

\vskip 0.75cm
\begin{center}
{\Large\bf 
Non-perturbative renormalization of the static\\[0.5ex] 
axial current in quenched QCD
}
\end{center}
\vskip 1.0cm
\vbox{
\centerline{
\epsfxsize=2.5 true cm
\epsfbox{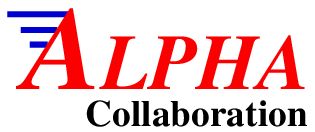}}
}
\vskip 1.0cm
\begin{center}
{\large
Jochen Heitger$^{\scriptscriptstyle a}$,
Martin Kurth$^{\scriptscriptstyle b}$ and
Rainer Sommer$^{\scriptscriptstyle c}$
}
\vskip 1.0cm
$^{\scriptstyle a}$
Westf\"alische Wilhelms-Universit\"at M\"unster,
Institut f\"ur Theoretische Physik, \\
Wilhelm-Klemm-Strasse~9, D-48149 M\"unster, Germany
\vskip 2.5ex
$^{\scriptstyle b}$
University of Southampton,
Department of Physics and Astronomy, \\
Highfield, Southampton SO17~1BJ, United Kingdom
\vskip 2.5ex
$^{\scriptstyle c}$
Deutsches Elektronen-Synchrotron DESY, Zeuthen, \\
Platanenallee~6, D-15738 Zeuthen, Germany
\vskip 0.875cm
{\bf Abstract}
\vskip 0.7ex
\end{center}

We non-perturbatively calculate the scale dependence of the static axial 
current in the Schr\"odinger functional scheme by means of a recursive 
finite-size scaling technique, taking the continuum limit in each step. 
The bare current in the $\Or(a)$ improved theory as well as in the 
original Wilson regularization is thus connected to the renormalization 
group invariant one. 
The latter may then be related to the current at the B-scale defined such 
that its matrix elements differ from the physical (QCD) ones by 
$\Or(1/M)$. 
At present, a (probably small) perturbative uncertainty enters in this 
step.
As an application, we renormalize existing unimproved data on $\Fbbare$
and extrapolate to the continuum limit.
We also study an interesting function $h(d/L,u)$ derived from the 
Schr\"odinger functional amplitude describing the propagation of a static 
quark-antiquark pair.



\vskip 0.29cm
\vfill

\begin{center}
February 2003
\end{center}

\eject
\vfill
\eject

\end{titlepage}

%% file: sect1.tex
\section{Introduction}
Since the decay constant $\Fb$, governing the leptonic decay of a B-meson, is  
an essential element in the quantitative analysis of the unitarity triangle,
many lattice QCD investigations have worked towards its
determination. However, with its large mass, the b-quark still escapes 
a direct numerical treatment \cite{lat02:bphys} and effective theories have 
to be used to separate the large mass scale from the low-energy bound-state 
dynamics. 
(As an exception to this rule, it has recently been demonstrated that also 
finite-volume methods on lattices with a large number of points represent a 
possible alternative \cite{fb:roma2a,fb:roma2b}.)

The most natural and theoretically appealing 
effective theory is the \emph{static approximation}~\cite{stat:eichhill1}.
It describes the large-mass limit of the theory and is
the starting point for a $1/m$--expansion, the heavy quark effective theory. 
Yet the problems of this framework have been twofold in the past.
(i) The renormalization properties 
of the static theory are different, i.e.~the renormalization constant 
$\zastat$ of the axial current in
$\Arstat=\zastat(\mu)\,\bpsid\gamma_0\gamma_5\psib^{\rm stat}$ becomes 
scale ($\mu$) dependent, thereby entailing an additional uncertainty, 
and (ii) Monte Carlo computations of the matrix element itself are difficult.
For these reasons, after a significant initial effort
\cite{fbstat:old1,fbstat:old2,fbstat:old3,fbstat:old4,fbstat:old5,stat:fnal2,stat:DMcN94,fbstat:old6},
the computation of $\Fb$ in the static approximation has received little 
attention in the recent past.  

In the present work, we solve (i) by computing the renormalization factor
that relates the bare operator in lattice regularization 
to the \emph{renormalization group invariant (RGI) operator}.
Denoting its matrix element by $\FhatstatRGI$, one then has a relation
\be \label{e_FB}
   \Fb \sqrt{\mB} = \Cps(\Mb/\Lambda_\msbar) \times \FhatstatRGI 
                    + \rmO(1/M_{\rm b})
\ee
with a function $\Cps$ of the renormalization group invariant b-quark mass 
$\Mb$ in units of the $\Lambda$--parameter.
It is scale independent, but in practice it is obtained perturbatively and 
an uncertainty due to perturbation theory remains, see \Sect{Sec_MEfinite}.
The important task of a lattice 
computation of the B-meson decay constant in the static approximation
is to compute $\FhatstatRGI$.

Our strategy to arrive at $\FhatstatRGI$ from the bare matrix element
follows the one used  by the ALPHA Collaboration for the 
renormalization group invariant quark mass \cite{msbar:pap1}. 
In this approach, an intermediate finite-volume
renormalization scheme is used to follow the scale evolution 
non-perturbatively to high energies ($\rmO(100 \GeV)$), where then 
perturbation theory can safely be used to connect to the 
renormalization group invariants.
For a more detailed explanation of the overall strategy we refer 
to \Ref{lat02:rainer} and for a preliminary report on our work 
to \Refs{ichep00:jochen,lat02:zastat}. 

The present paper is organized as follows.
In \Sect{Sec_rscheme} we discuss our intermediate renormalization scheme,
formulated in the Schr\"odinger functional (SF). It is based
on the renormalization condition introduced in \cite{zastat:pap1},
but a modification has been necessary to achieve
good statistical precision. \Sect{Sec_NPrun} contains the numerical 
determination of the scale dependence of the current in the SF scheme which 
is independent of the lattice formulation. We also relate
the current renormalized at some proper low scale to the RGI current. 
\Sect{Sec_match} gives our results for the lattice formulation dependent 
values of the $Z$--factor at this low scale. 
In \Sect{Sec_MEfinite} we then discuss \Eq{e_FB}
and explain in detail how our results are to be used. As an example we
obtain $\Fbsstat$ from published numbers of the bare
matrix element.  We finish 
with a brief discussion of the results in \Sect{Sec_discu}. 
Details of the numerical and perturbative calculations are described in 
appendices.

%% file: sect2.tex
\section{Intermediate renormalization scheme}
\label{Sec_rscheme}
In this section we introduce our intermediate renormalization scheme.
For reasons to be explained below, it differs from the
one originally introduced in \Ref{zastat:pap1}.
The perturbative calculations of \Ref{zastat:pap1}
are extended to the new scheme in \APP{App_pert}.

We choose a mass-independent renormalization scheme, leading
to simple renormalization group equations.
The scheme is defined using the Schr\"odinger functional 
(SF) \cite{SF:LNWW,SF:stefan1}, i.e.~the QCD partition function
$\mathcal{Z}
=\int_{T\times L^3}\rmD[A,\psibar,\psi]\,\rme^{-S[A,\psibar,\psi]}$
on a $T\times L^3$ cylinder in Euclidean space, where periodic boundary 
conditions in the spatial directions of length $L$ and Dirichlet boundary 
conditions at times $x_0=0,T$ are imposed on the gluon and quark 
fields.\footnote{
The spatial boundary conditions of the quark fields are only periodic
up to a global phase $\theta$ \cite{pert:1loop}, an additional 
`kinematical' parameter.}
Their exact form can be found in \Ref{impr:pap1}.
Moreover, we set $T=L$ throughout, and the renormalization scale $\mu$ is 
identified with the inverse box extension, $1/L$.
Such a finite-volume renormalization scheme is chosen,
since it allows to study the scale dependence in the 
{\em continuum limit} for a large range of $\mu$
\cite{alpha:sigma,alpha:SU2,alpha:SU3,msbar:pap1}. We can then
relate the quantities renormalized at some low scale
$\mu$ to the RGI quantities. 
Reviews of the strategy are found in
\cite{schlad:rainer,reviews:leshouches,lat02:rainer}, 
and for a more detailed description the reader should 
consult \Ref{msbar:pap1} which we will follow quite closely. 

As detailed in \cite{zastat:pap1},
we consider the SF with vanishing  
boundary gauge fields and $\theta=0.5$.
These settings are identical to those used for the quark mass 
renormalization in \cite{msbar:pap1} and were motivated by meeting the
criteria of good statistical precision of the Monte Carlo results, 
well-behaved perturbative expansions of the renormalization group
functions and minimization of lattice artifacts \cite{mr:pert}.
Static quarks are included as discussed in \Ref{zastat:pap1}, and we
use the notation of that reference. Throughout most of this paper, we
formally stay in the framework of continuum QCD; some notation and 
basic formulae of the lattice regularized theory, in which the following
expressions receive a precise meaning, are collected in \APP{App_latSSF}.

In contrast to the relativistic current, there is no axial Ward identity 
which protects the renormalized static-light axial vector current,
\be
\Aren(x)=
\zastat\,\lightb(x)\gamma_0\gamma_5\heavy(x) \,,
\ee
from a scale dependence.
Its scale evolution is 
governed by the renormalization group equation
\be
\mu\,{{\partial\,\Phi}\over{\partial\mu}}=\gamma(\gbar)\Phi \,,
\ee
where 
\be
\Phi\equiv\Fhatstat=\ketbra{\,{\rm f}\,}{\Aren}{\,{\rm i}\,}
\label{me_astat}
\ee
is an arbitrary matrix element of the renormalized static current. 
The renormalization group function $\gamma$, the anomalous dimension, has 
a perturbative expansion
\be
\gamma(\gbar)
\stackrel{\gbar\rightarrow0}{\sim}
-\gbsq\,\Big\{\,
\gamma_0+\gamma_1\gbsq+\gamma_2\gbar^4+\Or(\gbar^6)\,\Big\} \,
\ee
with a universal leading coefficient \cite{Shifman:1987sm,Politzer:1988wp},
\be
\gamma_0= 
-{1\over{4\pi^2}} \,,
\label{gam_0}
\ee
and $\gamma_1,\gamma_2,\ldots$ depending on the chosen renormalization
scheme.

Non-perturbatively, one computes the change of $\Phi$
under finite changes of the renormalization scale. For a scale factor
of two, the induced change defines the \emph{step scaling function},
\be
  \sigmaAstat(u) = \Phi(\mu/2)/\Phi(\mu) = \zastat(2L) / \zastat(L)\,,
\label{SSF_cont}
\ee
whose argument $u\equiv\gbar^2(L)$ is taken to be exactly the coupling 
defined in \cite{alpha:SU3}, and as always in the SF we have $\mu=1/L$.

\subsection{The old scheme}
\label{Sec_rscheme_old}
\begin{figure}[htb]
\centering
\epsfig{file=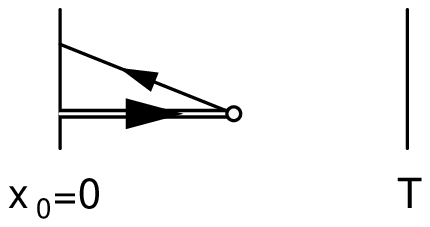,width=6.0cm}
\epsfig{file=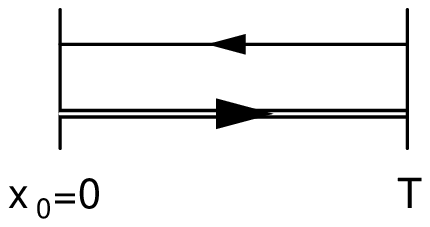,width=6.0cm}
\vspace{-0.5cm}
\caption{{\sl 
Sketch of the correlation functions $\fastat(x_0)$ (left) and 
$\fonestat$ (right).
The single and double lines represent light (i.e.~relativistic) and static
quark propagators, respectively. 
}}\label{fig:fa+f1_stat}
\end{figure}

In \Ref{zastat:pap1}, a normalization condition was 
formulated in terms of suitable correlation functions 
defined in the SF. It reads
\be
  \zastat(L)\, X(L) = X^{(0)}(L) \quad\mbox{at vanishing quark mass} \,,
\label{rencond_old}
\ee
with 
\be
X(L) = {\fastat(L/2)\over\sqrt{\fonestat}}\,
\label{X_old}
\ee
and $X^{(0)}(L)$ the tree-level value of $X(L)$.
Here, $\fastat$ is a correlation function between a static-light 
pseudoscalar boundary source and $\Astat$ in the bulk, and $\fonestat$ 
denotes the correlator between two such boundary sources at $x_0=0$ and 
$x_0=T$:
\bea
\fastat(x_0)
& = &
-{1\over 2}\int\rmd^3\vecy\,\rmd^3\vecz\,
\mvl{\Astat(x)\,\zetahb(\vecy)\gamma_5\zetal(\vecz)} \,,\\
\fonestat
& = &
-{1\over{2L^6}}\int\rmd^3\vecu\,\rmd^3\vecv\,\rmd^3\vecy\,\rmd^3\vecz\,
\mvl{\zetalbprime(\vecu)\gamma_5\zetahprime(\vecv)\,
\zetahb(\vecy)\gamma_5\zetal(\vecz)} \,.
\eea
(For the proper definition of the `boundary quark and antiquark
fields' $\zeta,\overline{\zeta}$ we refer to 
\Refs{zastat:pap1,impr:pap1}.)
The two correlators are schematically depicted in \Fig{fig:fa+f1_stat}, 
and their explicit form on the lattice is given in Eqs.~(\ref{fastat}) 
and (\ref{f1hl}) of \App{App_latSSF_defs}.
%
%
In the ratio (\ref{X_old}) both the multiplicative renormalization of the 
boundary quark fields and the  mass counterterm of the static field cancel.

\begin{figure}[htb]
\centering
\vspace{-1.5cm}
\epsfig{file=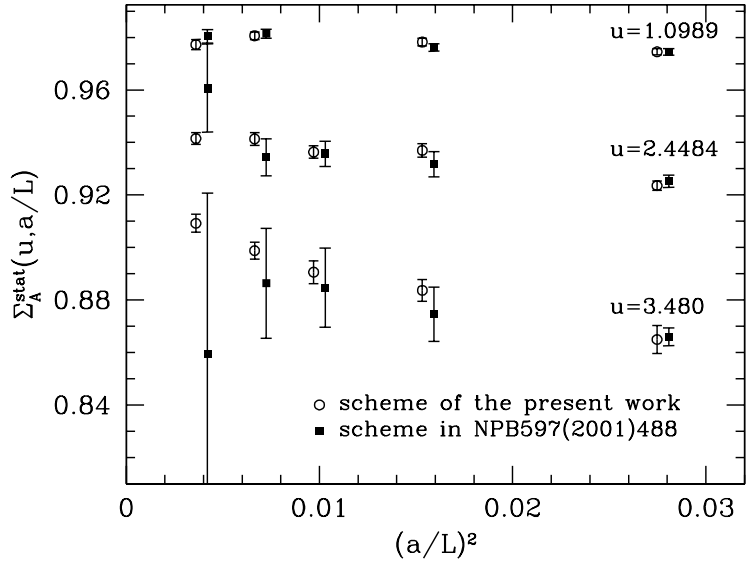,width=11.0cm}
\vspace{-1.75cm}
\caption{{\sl 
Comparison of the numerical precision of the lattice step scaling function
computed in the scheme of \protect\Ref{zastat:pap1} with the new one 
introduced in Subsection~\ref{Sec_rscheme_new}, which will be used in the 
rest of this paper. 
(The symbols are slightly displaced for better readability.)
The statistics of both computations is the same.
}}\label{fig:Sigma_zastat12}
\end{figure}

We now have to point out a drawback of this scheme 
that only becomes evident, when it is implemented 
numerically.
Namely, the lattice step scaling function $\SigmaAstat(u,a/L)$ 
(cf.~\Eq{SSF_lat} for its definition), computed by means of Monte Carlo 
simulations, has large statistical errors at $u\approx1.5$ and larger.  
In particular, these errors grow with $L/a$.
This can be inferred from the results tabulated 
in \App{App_latSSF_res} and
is illustrated for three representative coupling values in 
\Fig{fig:Sigma_zastat12}.
%
%
For $L/a=12,16$ (which amounts to also calculate $\zastat$ 
for $2L/a=24,32$), already around
$u\approx1.5$ a precise determination of $\SigmaAstat(u,a/L)$ 
with a reasonable computational effort becomes impossible.
The reason for this lies in the boundary correlator $\fonestat$
being part of the renormalization condition \Eq{rencond_old}: 
it contains the static quark propagating over a distance $T=L$.  
Thus $\fonestat$ falls roughly like $\rme^{\,-e_1 g_0^2 \, T/a}$ with
$e_1=\frac{1}{12\pi^2}\times19.95$ \cite{stat:eichhill_za} the leading 
coefficient of the self-energy of a static quark. 
On the other hand, the statistical errors fall much
more slowly, leaving us with an exponential 
degradation of the signal-to-noise ratio.

To circumvent this problem, we now introduce a slightly modified 
renormalization scheme. 
(Therefore, for the rest of the paper, the scheme of \Ref{zastat:pap1} 
--- if mentioned at all --- will only be referred to by labelling the 
corresponding quantities with an additional `$\oldSF$',
e.g.~$\zastat\rightarrow\zastatold$.) 
The general idea is to replace $\fonestat$ containing a static and a light 
quark by two boundary-to-boundary correlation functions. 
One of them contains a light quark-antiquark pair, the other
a static quark-antiquark pair.
Both can be computed with small
statistical errors, the latter because the variance reduction method
of \Ref{PPR} can be applied. Since the static-static boundary correlator
has not been studied before, we discuss it in some detail.
\subsection{The static-static boundary correlator $\fonehh$}
\label{Sec_rscheme_f1hh}
\begin{figure}[htb]
\centering
\epsfig{file=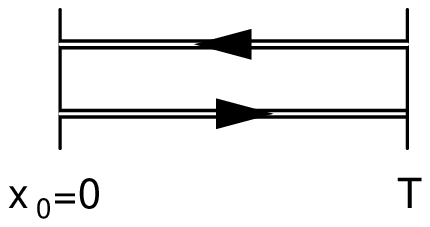,width=6.0cm}
\epsfig{file=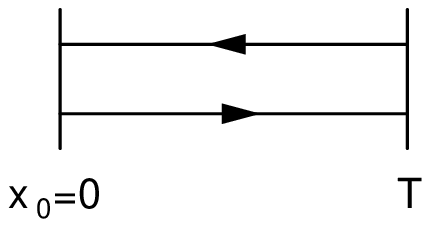,width=6.0cm}
\vspace{-0.5cm}
\caption{{\sl 
The correlation functions $\fonehh(x_3)$ (left) and $\fone$ (right).
The notation is the same as in \Fig{fig:fa+f1_stat}.
}}\label{fig:fahh+f1}
\end{figure}

We define
\be
\fonehh(x_3)\equiv
-{1\over{2L^2}}\int\rmd x_1\,\rmd x_2\,\rmd^3\vecy\,\rmd^3\vecz\,
\mvl{\zetahbprime(\vecx)\gamma_5\zetahprime({\bf 0})\,
\zetahb(\vecy)\gamma_5\zetah(\vecz)} \,,
\ee
represented graphically in
the left part of \Fig{fig:fahh+f1}.
%
%
After integrating out the static quark fields, 
$\fonehh$ becomes the (trace of the) product of a timelike
Wilson line and the complex conjugate one 
from boundary to boundary (see \Eq{e_fonehh_app}). 
They are separated by $\vecx$ in space.
This quantity is integrated over $x_1,x_2$ but retains a dependence on $x_3$.
In the following we take $d\equiv|x_3|$ as its argument, where 
the periodicity of the system in the space directions restricts
it to $0 \le d \le L/2$.

Upon renormalization the correlation function $\fonehh$ becomes
\bean
(\fonehh)_{\rm R}(d)=
\rme^{\,-2\,\delta m\times L}(\zh)^4\fonehh(d) \,,
\eean
where $\zh$ is the wave function renormalization
constant of a static boundary quark field and $\delta m$ the linearly
divergent static mass counterterm.
Therefore, to study the properties of $\fonehh$ further, we form the finite 
ratio
\be \label{e_h}
h(d/L,u)\equiv
{(\fonehh)_{\rm R}(d)\over(\fonehh)_{\rm R}(L/2)}=
{\fonehh(d)\over\fonehh(L/2)}  
\quad\mbox{at}\quad \gbar^2(L)=u \,.
\ee
Considered on the lattice, it has a continuum limit. 
As outlined in \App{App_pert_f1hh},
the one-loop coefficient $h^{(1)}(d/L)$ of the 
perturbative expansion
\be \label{e_hpert}
 h(d/L,u) = 1 + u\,h^{(1)}(d/L) + u^2 h^{(2)}(d/L) + \ldots
\ee
is given by
\be \label{e_honeloop}
 h^{(1)}(d/L) =  {2\over3}\left({1\over 2}-{d \over L}\right)^2\,.
\ee
Remarkably, this form holds exactly on the lattice
without any $a/L$--dependence. Some insight why this is so is presented in
the appendix as well.
At two-loop accuracy, we do not
expect exact $a$--independence any more, but still one may hope
that the favourable kinematics keep lattice artifacts small. 

\begin{figure}[htb]
\centering
\vspace{-1.5cm}
\epsfig{file=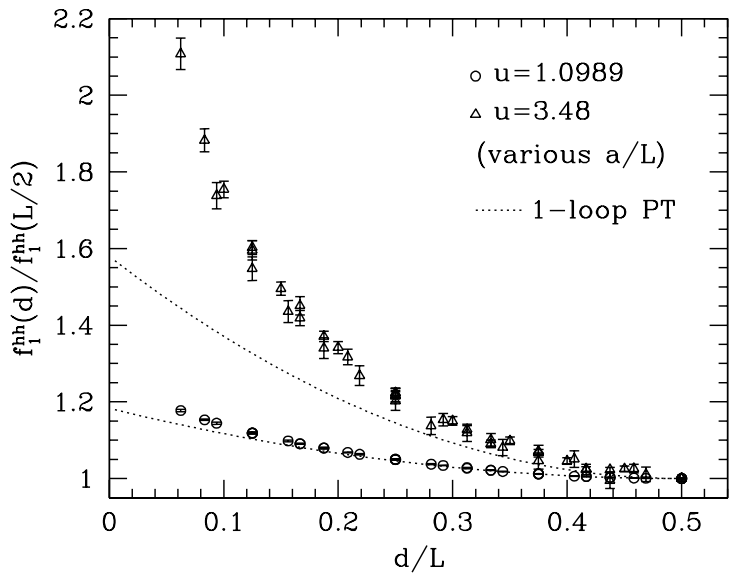,width=11.0cm}
\vspace{-1.75cm}
\caption{{\sl
Non-perturbative $d$--dependence of $h(d/L,u)=\fonehh(d)/\fonehh(L/2)$ at 
two (low and high) values of the coupling, compared to one-loop 
perturbation theory.
As data points corresponding to various lattice resolutions are included,
the figure also reflects the weak cutoff dependence of this ratio.
}}\label{fig:f1hh_d}
\end{figure}

The one-loop expression is compared to results from our non-perturbative 
computation of $\fonehh$ for two representative values of the coupling 
in \Fig{fig:f1hh_d}.
%
%
The figure contains non-perturbative results for 
$L/a \in \{12,16,20,24,32\}$ but at the level of our statistical errors, 
which are about 1\% and smaller, no lattice artifacts of the ratio $h$ can 
be seen. 

For low $d$, the non-perturbative data for $h$ are well described 
by $c\times(L/d)+c\,'$, where the
constant $c$ grows with $u=\gbar^2$. Hence the correlation
function contains a non-integrable short-distance singularity, which 
is the reason why we will not integrate over $d$ in the following. 
It is easy to see that this singularity is absent
up to and including 
the order $u^2$,  but in higher-order terms in 
perturbation theory such a singularity may appear.
\subsection{The new renormalization scheme}
\label{Sec_rscheme_new}
Choosing $d$ at its maximum value to further keep discretization errors
at a minimum, we specify our (non-perturbative) renormalization scheme by
\be
  \zastat(L)\, \Xi(L) = \Xi^{(0)}(L) \quad\mbox{at vanishing quark mass} \,,
\label{rencond}
\ee
with 
\be
\Xi(L) = {{\fastat(L/2)}\over\left[\,f_1\,\fonehh(L/2)\,\right]^{1/4}}\,.
\label{X_new}
\ee
Here, $\fone$ is the correlator between two light-quark pseudoscalar boundary 
sources,
\be
\fone=
-{1\over{2L^6}}\int\rmd^3\vecu\,\rmd^3\vecv\,\rmd^3\vecy\,\rmd^3\vecz\,
\mvl{\zetalbprimeone(\vecu)\gamma_5\zetalprimetwo(\vecv)\,
\zetalbtwo(\vecy)\gamma_5\zetalone(\vecz)} \,,
\ee
depicted in the right part of \Fig{fig:fahh+f1}.
The form of $\fone$ and $\fonehh$ on the lattice is given in
Eqs.~(\ref{f1}) and (\ref{f1hh}) of \App{App_latSSF_defs}.
As before, the combination of these correlators in the denominator of
(\ref{X_new}) is such that the boundary field renormalizations and the
mass counterterm drop out and no other scale but $L$ appears.

For $\theta=0.5$ the perturbative calculation summarized in 
\APP{App_pert} now yields
\be
\gamSF_1=
{1\over{(4\pi)^2}}\,\left\{\,0.10(2)-0.0477(13)N_{\rm f}\,\right\}\,,
\label{gamSF_1}
\ee
which differs only little from the one in the old scheme~\cite{zastat:pap1}.

Note that $\Oa$ improvement~\cite{impr:sym1,impr:pap1} 
can be applied and is an important 
ingredient in practice to reduce the cutoff effects in the numerical 
simulations (see \APP{App_latSSF}). Returning to \Fig{fig:Sigma_zastat12},
one observes that the statistical errors of the lattice results 
are indeed much smaller in the new scheme.

%% file: sect3.tex
\section{Non-perturbative running and renormalization group invariant} 
\label{Sec_NPrun}
In this section we present our quenched
results on the evolution of $\Phi(\mu)$  
over more than
two orders of magnitude in $\mu$.
To this end we consider the evolution of $\zastat$ under repeated changes 
of the scale (i.e.~the box size $L$) by a factor of two at fixed bare 
parameters.
Starting at some initial low-energy value (i.e.~some large $L=\Lmax$), 
one thereby climbs up the energy scale by repeated application of the 
inverse of the step scaling function until the perturbative domain at high 
energies (i.e.~small $2^{-k}\Lmax$) is reached, where finally the 
associated (scale and scheme independent) renormalization group invariant 
may be extracted.
As in the previous section we keep the discussion in the continuum theory
here; the underlying lattice calculations are described 
in \APP{App_latSSF}.
\subsection{Step scaling function}
%
%
\begin{figure}[htb]
\centering
\vspace{-1.5cm}
\epsfig{file=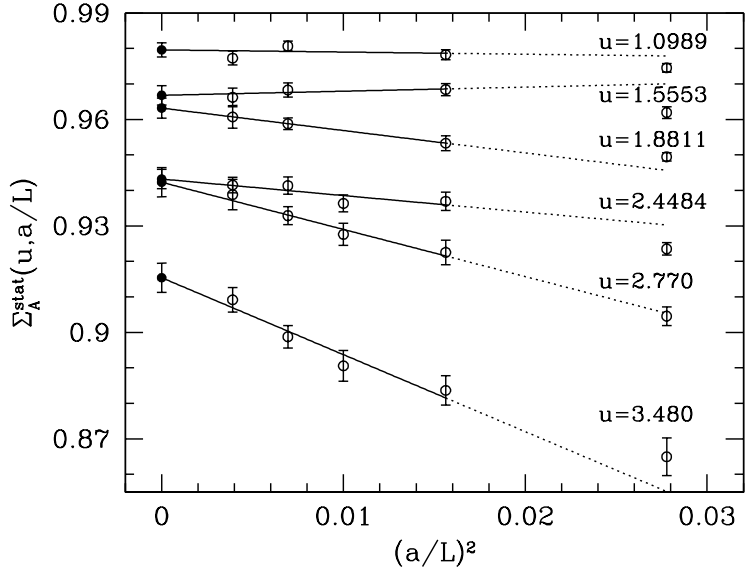,width=11.0cm}
\vspace{-1.75cm}
\caption{{\sl
Lattice step scaling function $\SigmaAstat$ and its continuum limit 
extrapolations for some selected values of $u$.
}}\label{fig:Sigma_zastat1}
\end{figure}
The evolution of $\zastat$ from size $L$ to $2L$ is given by 
its step scaling function, $\sigmaAstat(u)$, which has already
been introduced in \Eq{SSF_cont}, but where it is understood that
$\zastat$ is defined in the new renormalization scheme according to
\Eq{rencond}. 

As detailed in \App{App_latSSF_res}, the sets of lattice parameters
$(L/a,\beta,\kappa)$, which in practice are required to non-perturbatively 
compute $\sigmaAstat(u)$, can be taken over from the   
quark mass renormalization \cite{msbar:pap1}.
The available coupling values $u$ allow to trace the scale dependence of
$\zastat$ up to $L=2\Lmax$, where the scale $\Lmax$ is implicitly defined
through
\be
\gbsq(\Lmax)=3.48 \,.
\label{gbar_lmax}
\ee
The sequence 
\be
  u_k = \gbar^2(2^{-k}\Lmax)\,,\quad k=0,\ldots,8\,,
\ee
is known from \Ref{msbar:pap1}, and thus the corresponding
sequence 
\be
v_k\equiv{{\zastat(2^{-k+1}\Lmax)}\over{\zastat(2\Lmax)}}
= {{\Phi(2^{k-1}/\Lmax)}\over{\Phi((2\Lmax)^{-1})}}\,, \quad v_0=1\,,
\ee
is simply given by
\be
v_0=1\,,\quad
v_{k+1}={{v_k}\over{\sigmaAstat(u_k)}}\,,
\label{recursion}
\ee
once the function $\sigmaAstat(u)$ is available in the corresponding
range of $u$.

%
%
\begin{table}[htb]
\centering
\vspace{0.25cm}
\begin{tabular}{ccccc}
\hline \\[-2.0ex]
  $u$ &&& $\sigmaAstat(u)$ \\[1.0ex]
\hline \\[-2.0ex]
  1.0989 &&& 0.9796(20)    \\
  1.3293 &&& 0.9746(25)    \\
  1.4300 &&& 0.9719(26)    \\
  1.5553 &&& 0.9668(27)    \\
  1.6950 &&& 0.9727(28)    \\
  1.8811 &&& 0.9632(28)    \\
  2.1000 &&& 0.9589(35)    \\
  2.4484 &&& 0.9432(27)    \\
  2.7700 &&& 0.9423(41)    \\
  3.4800 &&& 0.9154(41)    \\[1.0ex]
\hline \\[-2.0ex]
\end{tabular}
\caption{{\sl 
Results for the continuum step scaling function 
$\sigmaAstat(u)$.
}}\label{tab:sigastat_res}
\end{table}
%
%
\begin{figure}[htb]
\centering
\vspace{-2.0cm}
\epsfig{file=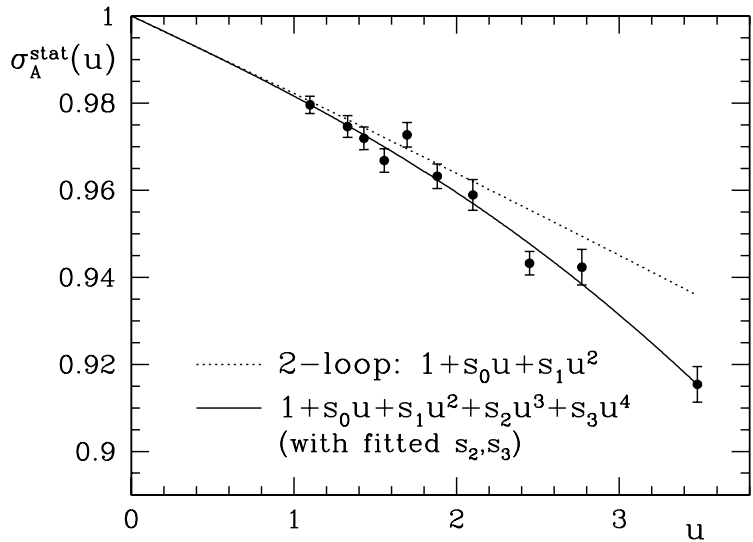,width=11.0cm}
\vspace{-2.25cm}
\caption{{\sl
Continuum step scaling function $\sigmaAstat(u)$ and its polynomial fit.
}}\label{fig:sigma_zastat1}
\end{figure}
The calculation of the lattice step scaling function and its
subsequent continuum extrapolation  
yields the pairs $u$ and $\sigmaAstat(u)$ listed 
in \Tab{tab:sigastat_res}. An impression of the quality of the 
continuum extrapolation is gained from \Fig{fig:Sigma_zastat1},
but for a more detailed account of the lattice
simulations and data analysis we refer to \App{App_latSSF_res}. 
An interpolating fit of $\sigmaAstat(u)$ is shown 
in \Fig{fig:sigma_zastat1}. The leading coefficients 
($s_0$ and $s_1$, see \App{App_contSSF_err}) of the 
interpolating polynomial are fixed to the 
perturbative predictions, \Eqs{e_s01}.
This fit is then inserted into the aforementioned recursion, and
propagating all errors through the recursion we obtain 
\be
\zastat(2\Lmax)/\zastat(L)=0.7551(47) 
\quad \mbox{at} \quad L=2^{-6}\Lmax \,,
\label{Zratio_res}
\ee
with the value of the coupling at this box size being 
$\gbsq(2^{-6}\Lmax)=1.053(12)$ \cite{msbar:pap1}.
Let us emphasize once more that $L=2\Lmax$ and $L=2^{-6}\Lmax$ represent 
low- and high-energy scales, respectively, which in this way have been 
connected non-perturbatively. (Our data actually allow to go up 
to $L=2^{-8}\Lmax$.)
\subsection{RGI matrix elements of the static axial current}
We now proceed to relate the renormalized matrix element 
\be
\Phi(\mu)=\zastat(L)\Phibare(g_0) \,, \quad \mu=1/L \,,
\label{me_ren}
\ee
at $L=2\Lmax$ to the renormalization group invariant one
defined by\footnote{
In a loose notation, we take sometimes $L$ and sometimes $\mu=1/L$ as the 
argument of $\gbar$.}
\be
\PhiRGI= 
\Phi(\mu)\,\times\,
\left[\,2b_0\gbar^2(\mu)\,\right]^{-\gamma_0/2b_0}
\exp\left\{-\int_0^{\gbar(\mu)} \rmd g 
\left[\,{\gamma(g)\over\beta(g)}-{\gamma_0 \over b_0 g}\,\right]\right\}\,,
\label{me_RGI}
\ee
with the universal 
leading-order coefficients $b_0=11/(4\pi)^2$ and $\gamma_0=-1/(4\pi^2)$ of 
the $\beta$-- and $\gamma$--functions, respectively.
Casting this equation in the form
\bean
{\PhiRGI\over\Phi((2\Lmax)^{-1})}\,
& = &
{\zastat(1/\mu)\over\zastat(2\Lmax)} \nonumber\\
&   &
\times\,\left[\,2b_0\gbar^2(\mu)\,\right]^{-\gamma_0/2b_0}
\exp\left\{-\int_0^{\gbar(\mu)} \rmd g 
\left[\,{\gamma(g)\over\beta(g)}-{\gamma_0 \over b_0 g}\,\right]\right\}\,,
\eean
with $\mu=2^6/\Lmax$,
we see that the first factor is known from \Eq{Zratio_res},
while in the second one only couplings $\gbar^2 \leq 1.05$ contribute
and it can safely be evaluated by perturbation theory.
Still, for the perturbative error to be negligible, 
$\gamma$ has to be known to two-loop accuracy and $\beta$ to three-loop. 
Upon inserting $\gbsq(2^{6}/\Lmax)=1.053$ and numerical integration of the 
second factor we find
\be \label{e_ratio1}
\Phi(\mu)/\PhiRGI = 1.088(8)
\quad \mbox{at} \quad \mu=(2\Lmax)^{-1}
\ee
in the SF scheme. 
Entirely consistent numbers, with slightly larger errors, 
are obtained for $\Phi(\mu)/\PhiRGI$ if one switches to
perturbation theory at $\mu=2^{7}/\Lmax$ or $\mu=2^{8}/\Lmax$ 
instead. 

\begin{figure}[htb]
\centering
\vspace{-1.5cm}
\epsfig{file=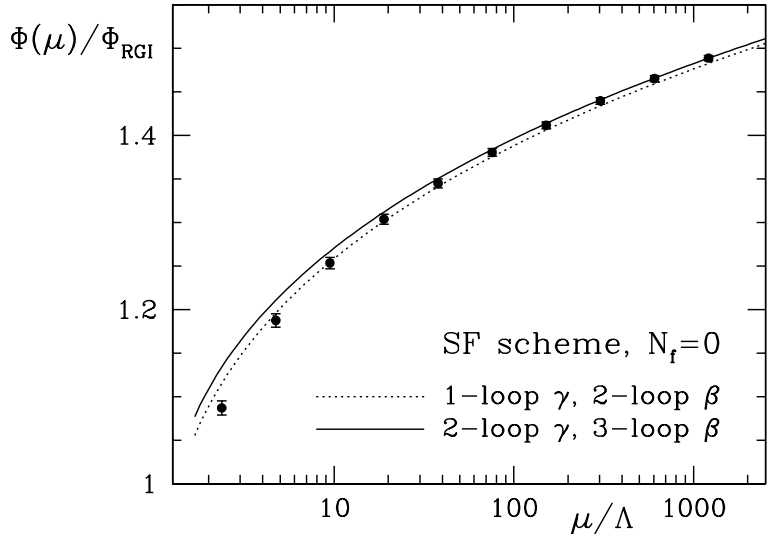,width=11.0cm}
\vspace{-2.0cm}
\caption{{\sl
Numerically computed values of the running matrix element of the static
axial current in the SF scheme compared to perturbation theory.
The dotted and solid lines are obtained from \Eq{me_RGI} using the
1/2-- and 2/3--loop expressions for the $\gamma$-- and $\beta$--functions,
respectively, as well as $\Lambda\Lmax=0.211$ from \protect\Ref{msbar:pap1}.
}}\label{fig:PhiSF_stat}
\end{figure}

In \Fig{fig:PhiSF_stat} we compare the numerically computed running 
with the corresponding curves in perturbation theory. While good
agreement with the perturbative approximation is seen at high scales, 
a growing 
difference of up to 5\% becomes visible when $\mu$ is lowered to 
$\mu\approx2.5\Lambda$.
%
%

Below it will be more convenient to specify the scale $\mu$ in \Eq{e_ratio1}
in terms of $r_0$ \cite{pot:r0} instead of $\Lmax$. 
Taking also  the small error contribution from the uncertainty
of $\Lmax$ in units of $r_0$, $\Lmax/r_0=0.718(16)$ \cite{pot:r0_ALPHA},
into account, the final result for the regularization independent part 
$\Phi(\mu)/\PhiRGI$ of the total renormalization factor is
\be
\Phi(\mu)/\PhiRGI = 1.088(10)
\quad{\rm at}\quad \mu=(1.436\,r_0)^{-1}\,.
\label{RGIfact_res}
\ee
Note that this result refers to the \emph{continuum limit} so that the 
error on $\Phi(\mu)/\PhiRGI$ of about 0.9\% should only be added in 
quadrature to the proper matrix element under study \emph{after} its 
continuum extrapolation.

%% file: sect4.tex
\section{$\zastat$ at low scale and total renormalization factor}
\label{Sec_match}
We still need to relate $\Arstat(\mu)$, renormalized at some appropriate
scale $\mu$, to the bare lattice operator.
This amounts to computing $\zastat$ at 
the low-energy matching scale $L=2\Lmax=1.436\,r_0$, which is briefly 
explained in \App{App_contSSF_match}. Since in this step the bare 
operator is involved, $\zastat$ does depend --- in contrast to the result 
of the previous section --- on the choice of action. We have considered 
three different cases. The first two are
the non-perturbatively $\Oa$ improved action
of \Ref{impr:pap3}, with $\castat=-\frac{1}{4\pi}\times g_0^2$ 
($=$ one-loop) and separately with $\castat=0$. Their combination will in 
the future allow to study the influence of $\castat$ on the
continuum extrapolations of renormalized matrix elements. The third 
choice is the unimproved Wilson action which is of interest,
because so far the best computations of the bare matrix element did not 
use improvement~\cite{stat:fnal2,stat:DMcN94}. 

\begin{figure}[htb]
\centering
\vspace{-1.5cm}
\epsfig{file=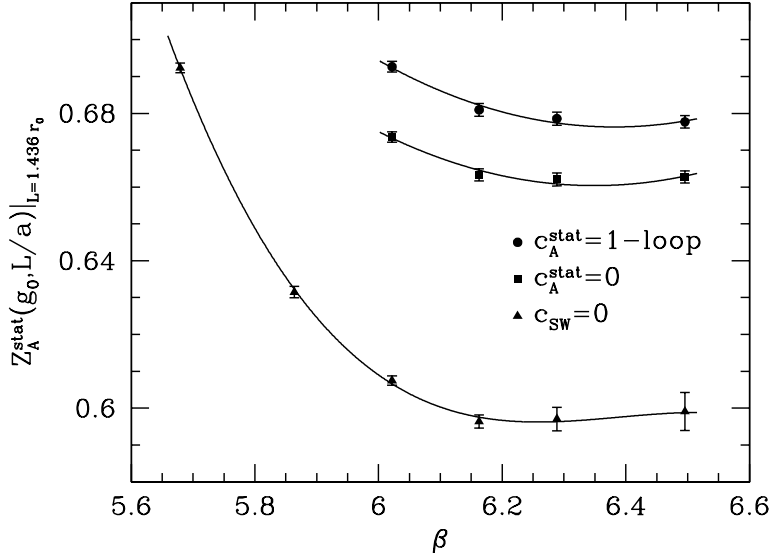,width=11.0cm}
\vspace{-1.75cm}
\caption{{\sl
Numerical results for $\zastat(g_0,L/a)\big|_{L=1.436\,r_0}$ together 
with their interpolating polynomials.}}\label{fig:zastat1_1.436r0}
\end{figure}

The numerical results for $\zastat$ are shown in \Fig{fig:zastat1_1.436r0}.
For later use they
are represented by interpolating polynomials,
\bea
\zastat(g_0,L/a)\,\Big|_{\,L=1.436\,r_0}=
\sum_{i\geq0} z_i \,(\beta-6)^i\,,
\label{e_ZApoly}
\eea
with coefficients $z_i$ as listed in \tab{tab:polys}.
The statistical uncertainty to be taken into account when using this
formula is about 0.4\%.
%
%
%
%
\begin{table}[htb]
\centering
\begin{tabular}{ccccr@{.}lr@{.}l}
\hline \\[-2.0ex]
  $\csw\,,\,\ca$ & $\castat$  & applicability 
& $i$ & \multicolumn{2}{c}{$z_i$} & \multicolumn{2}{c}{$f_i$} \\[1.0ex]
\hline \\[-2.0ex]
  non-perturbative  & $-g_0^2/(4\pi)$ & $6.0\leq \beta \leq 6.5$ 
& 0 & $0$&$6944$  & $0$&$6382$  \\
  \cite{impr:pap3}  &             &
& 1 & $-0$&$0946$ & $-0$&$0869$ \\
                    &             &
& 2 & $0$&$1239$  & $0$&$1139$  \\[1.0ex]
  non-perturbative  & $0$         & $6.0\leq \beta \leq 6.5$
& 0 & $0$&$6750$  & $0$&$6204$  \\
                    &             &
& 1 & $-0$&$0838$ & $-0$&$0771$ \\
                    &             &
& 2 & $0$&$1200$  & $0$&$1103$  \\[1.0ex]
   $0\,,\,0$        & $0$         & $5.7\leq \beta \leq 6.5$
& 0 & $0$&$6090$  & $0$&$5598$  \\
                    &             &
& 1 & $-0$&$1186$ & $-0$&$1090$ \\
                    &             &
& 2 & $0$&$3438$  & $0$&$3160$  \\
                    &             &
& 3 & $-0$&$2950$ & $-0$&$2711$ \\[1.0ex]
\hline \\[-2.0ex]
\end{tabular}
\caption{{\sl 
Coefficients of the interpolating polynomials, 
\Eq{e_ZApoly} and \Eq{e_Zphipoly}. 
Uncertainties are discussed in the text.
}}\label{tab:polys}
\end{table}
\subsection*{The total renormalization factor}
The total renormalization factor to directly translate any bare matrix 
element $\Phibare(g_0)$ of $\Astat$ into the RGI matrix element, 
$\PhiRGI$, can be written as
\be
\ZPhi(g_0)=
{\PhiRGI\over\Phi(\mu)}\,\bigg|_{\,\mu=(1.436\,r_0)^{-1}}
\,\times\,
\zastat(g_0,L/a)\,\Big|_{\,L=1.436\,r_0} \,.
\ee
We combine \Eq{e_ZApoly} with \Eq{RGIfact_res}
and represent the total $Z$--factor by further interpolating polynomials,
\bea
\ZPhi(g_0)=
\sum_{i\geq0} f_i \,(\beta-6)^i\,,
\label{e_Zphipoly}
\eea
whose coefficients are also found in \tab{tab:polys}.
These parametrizations of $\ZPhi$ are to be used with an uncertainty of 
about 0.4\%\footnote{
Only in the case $\csw=0$ the error to be associated with the formulae 
for $\zastat$ and $\ZPhi$ grows to 0.5\% at $\beta\approx 6.3$ and 0.8\% 
at $\beta\approx 6.5$.} 
at each $\beta$--value and an additional error of 0.9\% 
(from $\PhiRGI/\Phi(\mu)$), which remains to be added in quadrature 
{\em after} performing a continuum extrapolation.

%% file: sect5.tex
\section{Matrix elements at finite values of the quark mass} 
\label{Sec_MEfinite}
In order to use results from the static theory,
one still  has to relate its renormalization
group invariant matrix elements to those 
in QCD at finite values of the quark mass, $m$. This step may also be
seen as a translation to another scheme, defined by the condition
that matrix elements in the static effective theory renormalized in this 
scheme and at scale $\mu=m$ are the same as those in QCD up
to $1/m$--corrections. This scheme is therefore denoted as the
\emph{`matching scheme'} \cite{lat02:rainer}. 
Below, we will specify precisely which quark mass $m$ is to be taken.
\subsection{Conversion to the matching scheme}
\begin{figure}[htb]
\centering
\vspace{-0.5cm}
\epsfig{file=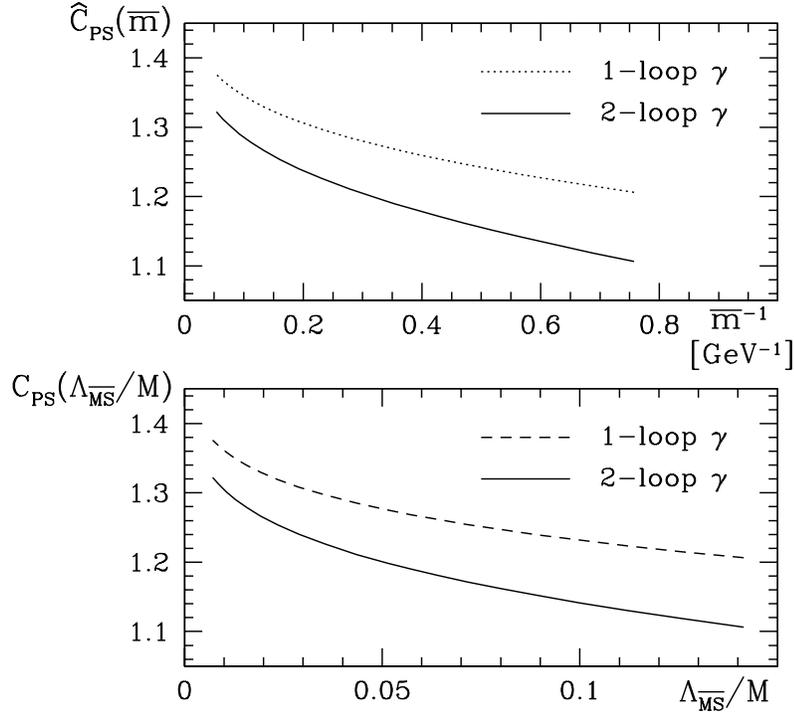,width=11.0cm}
\vspace{-1.0cm}
\caption{{\sl
Conversion factor to the matching scheme, which translates the RGI matrix 
element to the one at finite mass.
}}\label{fig:PhiMatch}
\end{figure}

Let us write the relations for the special case of the matrix element of the
axial current between the vacuum and the heavy-light pseudoscalar, 
\be
\PhiRGI=\ZPhi \ketbra{\,{\rm PS}\,}{\Astat}{\,{\rm 0}\,} \,.
\ee
We then have 
\be
   \Fp \sqrt{\mp} = \Cpshat(\mbar) \times \PhiRGI + \rmO(1/\mbar)\,,
 \label{e_Fpmp_msbar}
\ee
where $\mbar$ is the $\msbar$ quark mass at renormalization
scale $\mbar$.\footnote{
Note that in \cite{HQET:neubert} a similar equation 
with the pole mass instead of the $\msbar$ mass is written. 
At the two-loop order, which will be used below, this does formally 
not make any difference. However, the pole mass does not have a well-behaved
perturbative expansion \cite{renormalons}, and we therefore prefer 
a short-distance mass such as the $\msbar$ mass.}
The function $\Cpshat(\mu)$ is given by
\be
\Cpshat(\mu) = 
\left[\,2b_0\gbar^2(\mu)\,\right]^{\gamma_0/2b_0}
\exp\left\{\int_0^{\gbar(\mu)} \rmd g 
\left[\,{\gamma(g)\over\beta(g)}-{\gamma_0 \over b_0 g}\,\right]\right\}\,,
\label{fmbar}
\ee
with $\gbar(\mu)$ the $\msbar$ running coupling and $\gamma$ the anomalous
dimension in the matching scheme.
The latter is known to two loops  
\cite{Ji:1991pr,BroadhGrozin,Gimenez:1992bf,stat:eichhill_za} 
with $\gamma_0$ being the 
same as before and
\bes
\gamma_1
& \equiv & 
\gamma_1^{\rm match}=\gamma_1^\msbar-{b_0\over 3\pi^2} \,,\\
\label{e_gammamsbar}
\gamma_1^\msbar
&    =   &
-{1\over 576\pi^4}\left({127\over 2}+28\zeta(2)-5N_{\rm f}\right)\,.
\ees
For illustration, $\Cpshat(\mu)$ is plotted in the upper part 
of \Fig{fig:PhiMatch}, where for the numerical evaluation 
the $\beta$--function is always taken at 
four-loop precision \cite{vanRitbergen:1997va}, while to estimate the 
perturbative uncertainty we show the result for the one-loop and the 
two-loop approximation of $\gamma$. 
%
%

\Eq{fmbar} can be rewritten in a form  
displaying explicitly that also this step is not restricted to
perturbation theory. In terms of the renormalization group 
invariant quark mass, $M$, we have
\be
\Fp \sqrt{\mp} = \Cps(\M/\Lambda_\msbar) \times \PhiRGI 
                 + \rmO(1/M)\,,
\label{e_Fpmp_RGI}
\ee
where now only renormalization group invariants enter.
To evaluate $\Cps(\M/\Lambda_\msbar)$ in perturbative approximation, 
one changes the argument of the function $\Cpshat$, \Eq{fmbar}, 
by inserting
\be
M / \mbar(\mu) = 
\left[\,2b_0\gbar^2(\mu)\,\right]^{-d_0/2b_0}
\exp\left\{-\int_0^{\gbar(\mu)} \rmd g 
\left[\,{\tau(g)\over\beta(g)}-{d_0 \over b_0 g}\,\right]\right\}
\label{M_RGI}
\ee
together with the condition $\mbar(m_{\msbar})=m_{\msbar}$, where 
$\tau(\gbar)$ denotes as usual the renormalization group function of the 
renormalized (running) quark mass with universal leading-order coefficient
$d_0=8/(4\pi)^2$.
A numerical evaluation (with the four-loop 
$\tau$--function \cite{Chetyrkin:1997dh,Vermaseren:1997fq} in \Eq{M_RGI}) is 
shown in the lower part of \Fig{fig:PhiMatch}. \Eq{e_Fpmp_RGI} is not only 
the cleanest form from a theoretical point of view but it is also practical,
because the relation between bare quark masses in the $\Oa$ improved lattice 
theory and the renormalization group invariant mass $M$ is known 
non-perturbatively in the quenched approximation\cite{msbar:pap3,impr:babp}. 

For later convenience we represent $\Cps$ in terms of the variable 
$x\equiv 1/\ln\left(\M/\Lambda_{\msbar}\right)$ in a functional form 
motivated by \Eq{fmbar},
\be
\Cps=
x^{\gamma_0/2b_0}\left\{\,1-0.065\,x+0.048\,x^2\,\right\}\,, \quad
x= 1/\ln\left(\M/\Lambda_{\msbar}\right) \leq 0.52\,,
\label{fhat_poly}
\ee
with $b_0=11/(4\pi)^2$ and $\gamma_0=-1/(4\pi^2)$.
It describes the result for the two-loop approximation of the 
$\gamma$--function within less than 0.01\%. Of course in this step a
perturbative error is involved, which is difficult to estimate. 
Assuming a geometric growth of the coefficients
of the $\gamma$--function, we find that the $\gamma_2$--term
would cause a change by around 1\% at $\mbar=\mbMSbar$ 
and by 2.5\% at $\mbar=1.2\,\GeV$. Thus one may attribute
a 2--4\% error due to the perturbative approximation, which could be 
much reduced by a computation of the three-loop anomalous dimension.

In principle, $\Cps$ may be computed also 
non-perturbatively following the strategy outlined in \Ref{lat02:rainer}.
It is then defined only up to $1/M$--terms,
consistent with \Eq{e_Fpmp_RGI}.
\subsection{Application: first non-perturbative renormalization of 
$\Fbsstat$}
\label{Sec_MEfinite_appl}
We now take bare matrix elements of $\Astat$ for unimproved Wilson fermions
from the literature to obtain an estimate for $\Fbs$ in the static 
approximation. 
This exercise serves mainly to illustrate how to use our results.

{\bf 1.} The matrix elements are needed at a fixed
value of the light quark mass. To avoid issues in the extrapolation
to very light quarks, we here consider only $\Fbs$.
To fix the strange quark mass, we use that the sum of
the light quark masses is to a good approximation
proportional to the squared (light-light) pseudoscalar
masses, $\mp^2({\rm l},{\rm l})$ \cite{msbar:pap3}, and interpolate
the data for the decay constant of \cite{stat:fnal2,stat:DMcN94} 
as a function of $\mp^2({\rm l},{\rm l})r_0^2$ to
$\mp^2({\rm s},{\rm s})r_0^2=(2\mk^2-\mpi^2)r_0^2=
2\mk^2r_0^2/(1+m_{\rm l}/m_{\rm s})=3.0233$.  
(To arrive at the latter, we employed $\mk^2r_0^2=1.5736$ 
\cite{msbar:pap3} and $m_{\rm s}/m_{\rm l}=24.4$ from chiral perturbation
theory \cite{reviews:cpt_leutwyler}.)
The resulting dimensionless numbers $r_0^{3/2}\Phibare$ are listed 
in \Tab{tab:Phi_wil}.
%
%
\begin{table}[htb]
\centering
\begin{tabular}{cccr@{.}lr@{.}lr@{.}lr@{.}lr@{.}l}
\hline \\[-2.0ex]
  $\beta=6/g_0^2$ &&
& \multicolumn{2}{c}{5.7} & \multicolumn{2}{c}{5.9} 
& \multicolumn{2}{c}{6.0} & \multicolumn{2}{c}{6.1} 
& \multicolumn{2}{c}{6.3} \\[1.0ex]
\hline \\[-2.0ex]
  $r_0/a$             &&& 2&93(1)  & 4&48(2)  & 5&37(2)  & 6&32(3) 
& 8&49(4)  \\
  $r_0^{3/2}\Phibare$ &&& 4&75(25) & 4&09(21) & 3&94(13) & 3&79(36) 
& 4&00(29) \\
  $r_0^{3/2}\PhiRGI$  &&& 2&99(16) & 2&35(12) & 2&21(7)  & 2&09(20) 
& 2&20(16) \\[1.0ex]
\hline \\[-2.0ex]
\end{tabular}
\caption{{\sl
Matrix elements of $\Astat$ in units of $r_0$ for the light quark mass 
equal to the strange quark mass. Bare matrix elements come from
\protect\Ref{stat:fnal2}, with the exception of $\beta=6.0$ which is taken
from \protect\Ref{stat:DMcN94}.
The scale $r_0/a$ \protect\cite{pot:r0} is used as determined
in \protect\Ref{pot:r0_ALPHA}, and its uncertainty is already included 
here.
}}\label{tab:Phi_wil}
\end{table}

\begin{figure}[htb]
\centering
\vspace{-2.25cm}
\epsfig{file=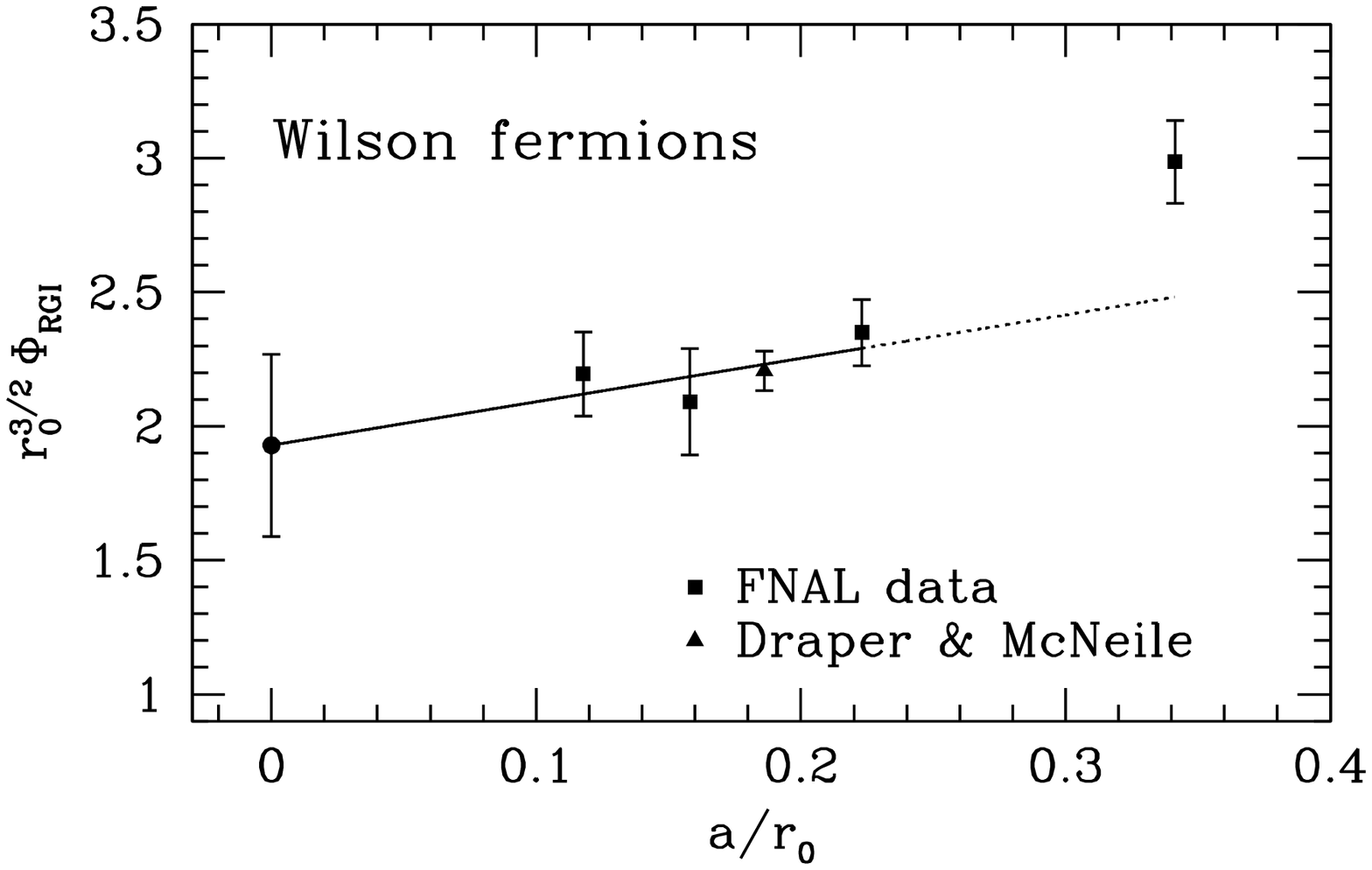,width=11.0cm}
\vspace{-2.5cm}
\caption{{\sl
Continuum extrapolation of the non-perturbatively renormalized matrix
element of $\Astat$ based on the unimproved Wilson data for $\Fbbare$ 
from \protect\Refs{stat:fnal2,stat:DMcN94}.
}}\label{fig:fBstat_wil}
\end{figure}

{\bf 2.} We renormalize by multiplying with \Eq{e_Zphipoly}, using the 
$f_i$ from \Tab{tab:polys} ($\csw=\ca=\castat=0$), take into account 
a 0.4--0.5\% error from the non-universal part of the $Z$--factor at each 
value of $g_0$ and find the last line in \Tab{tab:Phi_wil}. 
Assuming the leading linear behaviour in the lattice spacing to dominate 
for $a/r_0<1/4$, we extrapolate to the continuum limit as shown 
in \Fig{fig:fBstat_wil}. 
%
%
Adding to the extrapolation error in quadrature also the 0.9\% error
contribution of $\ZPhi$, which is independent of $g_0$, yields
\be
r_0^{3/2}\PhiRGI = 1.93(34) \quad\mbox{\rm at}\quad a=0 \,.
\label{e_res_Phi}
\ee

{\bf 3.} Finally, inserting $M_{\rm b}r_0=17.6(5)$ 
\cite{lat01:mbstat,lat02:rainer} and $\Lambda_\msbar r_0=0.602(48)$ 
\cite{msbar:pap1}, one gets via the formula in \Eq{fhat_poly} 
\be
\Cps(M_{\rm b}/\Lambda_\msbar) = 1.23(3) \,,
\ee
where a 2\% error for the perturbative approximation is assumed.
With the experimental spin-averaged B-meson mass 
$\mb=m_{{\rm B}_{\rm s}}=5.4\,\GeV$, we then obtain from \Eq{e_Fpmp_RGI}:
\bea
\Fbsstat r_0
& = & 
0.64(11) \,,\\
\Fbsstat
& = &
253(45)\,\MeV \quad\mbox{\rm for}\quad r_0=0.5\,\fm \,.
\eea
The result contains all errors apart from the uncertainty owing to the 
quenched approximation. Evidently, the continuum extrapolation may be done 
much better, once $\Oa$ improved results with sufficient precision and small 
lattice spacings are available.

%% file: sect6.tex
\section{Discussion}
\label{Sec_discu}
We have performed the scale dependent renormalization of $\Astat$ by
constructing a non-perturbative renormalization group in the \SF scheme, 
and agreement with perturbation theory at large scales was demonstrated.
The renormalization factors needed to extract the associated RG invariant 
are computed with good numerical accuracy, which is a crucial prerequisite 
for a controlled determination of $\Fb$ in the static limit.
In \Ref{lat02:rainer} it was shown that the renormalization factors
obtained in this way differ appreciably 
from earlier estimates \cite{stat:fnal2} 
based on tadpole-improved perturbation theory \cite{Lepenzie}. 
Hence their non-perturbative computation is important.

We have not emphasized this so far, but our computation
provides the scale dependence of {\em all} static-light 
bilinears 
\be
 \op{\Gamma}(x) = \lightb(x) \Gamma \heavy(x) \,.
\ee
They are renormalized by
\bea
(\op{\gamma_{\it k}})_{\rm R}(x) 
& = & 
\zastat \op{\gamma_{\it k}} \,, \nonumber \\
(\op{\gamma_0})_{\rm R}(x)       
& = & 
\Ztil\zastat \op{\gamma_0} \,,\quad 
(\op{\gamma_{\it k}\gamma_5})_{\rm R}(x)=
\Ztil\zastat\op{\gamma_{\it k}\gamma_5}\,,
\eea
with a scale independent renormalization $\Ztil$. This pattern is due to
the heavy quark spin symmetry which is exact on the lattice,
and due to the chiral symmetry of the continuum theory.  
The latter means that the relative renormalization $\Ztil$ 
may be fixed by imposing a suitable chiral Ward 
identity \cite{Zvstat:onogi} and is thus scale independent. 

Returning to the case of most interest, $\Fbstat$,
our continuum extrapolation in \Sect{Sec_MEfinite_appl} that uses 
unimproved data for the bare matrix elements from the literature 
and also quite large lattice spacings leaves much room for improvement of
the present result, $\Fbsstat=253(45)\,\MeV$. Apart from the obvious step
of obtaining $\Oa$ improved bare matrix elements at small lattice spacings
and extrapolating to the continuum, it will be necessary
to estimate the $\rmO(1/M)$ correction. 
There are two possible roads towards this goal. 

An elegant and clean way is to compute the $1/M$--corrections 
directly as perturbations to the static effective theory. Again, the
main problem here is renormalization. Indeed, this is a severe one,
since mixing between operators of different dimensions has to be
taken into account. This will require much more theoretical and
numerical effort; 
but a possible strategy exists \cite{lat01:mbstat,lat02:rainer}.

In the mean time, one may also compare the prediction from the static
approximation to what one obtains at $M\approx M_{\rm c}$
and also to the results obtained directly at $M=M_{\rm b}$,
most notably the ones of \Refs{fb:roma2a,fb:roma2b}. As emphasized in
\Refs{fb_wupp,reviews:beauty,reviews:beauty_hartmut} this should be
done in the continuum limit, since $\Oa$ errors get enhanced when
the quark masses increase. At the charm quark mass these are sizeable
but can be extrapolated away, at least in the quenched approximation 
\cite{mcbar:RS02}. The comparison between finite-mass decay constants
and $\Fbstat$ is most conveniently done by 
comparing $\Fp\sqrt{\mp}/\Cps(\M/\Lambda_\msbar)$.
Unfortunately, at present the error of the static result
is still too large to draw any strong conclusions about $1/M$--corrections
in $ F_{\rm B}$:
\bes
\label{e_fs}
r_0^{3/2}\,\PhiRGI 
& = &  
1.93(34)\,,\,\quad \mbox{static}:\,\;\mbox{\Eq{e_res_Phi}} \,,\\
\label{e_fb}
r_0^{3/2}\,{F_{\rm B_{\rm s}}\sqrt{m_{\rm B_{\rm s}}} 
\over\Cps(\Mbeauty/\Lambda_\msbar)} 
& = & 
1.46(23)\,,\,\quad \mbox{using} \quad F_{\rm B_{\rm s}}=192(30)\,\MeV\,\;
\mbox{\cite{fb:roma2a,fb:roma2b}} \,,\\
\label{e_fc}
r_0^{3/2}\,{F_{\rm D_{\rm s}}\sqrt{m_{\rm D_{\rm s}}} 
\over\Cps(\Mcharm/\Lambda_\msbar)} 
& = & 
1.29(\phantom{0}5)\,,\,\quad \mbox{using} \quad 
F_{\rm D_{\rm s}}=252(\phantom{0}9)\,\MeV\,\;
\mbox{\cite{fds:JR03}} \,.
\ees
Still, the difference of \Eq{e_fs} and \Eq{e_fc} shows that there are 
significant $1/M$--corrections in the charm mass region.

As a more technical remark we point out that the function $h(d/L,u)$, 
\Eq{e_h}, shows very small $a$--effects in the quenched approximation 
and may be worth studying to verify improvement with dynamical fermions.

\subsection*{Acknowledgements}
We thank DESY for allocating computer time on the APE-Quadrics computers
at DESY Zeuthen to this project.
This work is also supported in part by the EU IHP Network on 
\emph{Hadron Phenomenology from Lattice QCD} under grant HPRN-CT-2000-00145 
and by the DFG Sonderforschungsbereich SFB/TR~9.

%% file: apdxa.tex
\section{Computation of the lattice step scaling function}
\label{App_latSSF}
This appendix describes some details of the numerical simulations on the
lattice and the subsequent calculations that we have performed in order to 
determine the step scaling function for $\zastat$. 
At the beginning we also recall a few basic definitions and formulae, 
which are specific to the inclusion of static quarks and the correlation 
functions that are considered in the framework of the Schr\"odinger 
functional (SF).
As the impact of the static quarks on $\Or(a)$ improvement of the 
static-light sector has extensively been discussed 
in \Ref{zastat:pap1}, the reader might consult this reference for 
further details and any unexplained notation.
\subsection{Definitions}
\label{App_latSSF_defs}
\subsubsection*{Lattice action}
The total lattice action is given by the sum
\be
S[U,\lightb,\light,\heavyb,\heavy]=
\Sg[U]+\Sf[U,\lightb,\light]+S_{\rm h}[U,\heavyb,\heavy] \,,
\label{stotdef}
\ee
where $\Sg$ and $\Sf$ are the standard pure gauge and $\Oa$ improved Wilson
actions for relativistic (light) quarks, see Eqs.~(A.22) and 
(A.23) -- (A.26) of \Ref{zastat:pap1}, respectively, and $S_{\rm h}$ 
denotes the lattice action for the heavy quark:
\be
S_{\rm h}[U,\heavyb,\heavy]= 
a^4\sum_x\heavyb(x)\nabstar{0}\heavy(x) \,. 
\label{shdef}
\ee
The fields $\heavy$ and $\heavyb$ of the static effective theory are 
constrained in such a way 
(namely $P_+\heavy=\heavy$ and $\heavyb P_+=\heavyb$) that one is left with 
just two degrees of freedom per space-time point \cite{stat:eichhill1} and 
only the (time component of the) backward lattice derivative, 
$\nabstar{\mu}$, enters in the action (\ref{shdef}).
Hence, static quarks propagate only forward in time, which also reflects in 
the form of the associated quark propagator,
\bea
S_{\rm h}(x,y)
& = &
U(x-a\hat{0},0)^{-1}\,U(x-2a\hat{0},0)^{-1} \,\cdots\, U(y,0)^{-1} 
\nonumber \\
&   &
\times\,\theta(x_0-y_0)\delta(\vecx-\vecy)P_{+} \,,\quad
P_{\pm}=\hlf\,(1\pm\gamma_0) \,,
\eea
being just a straight timelike Wilson line.

To impose SF boundary conditions, \Eqs{stotdef} and 
(\ref{shdef}) are supplemented by 
\bea
\light(x)=0
& \quad &
\hbox{if $x_0<0$ or $x_0>L$} \,, \nonumber \\
\heavy(x)=0
& \quad &
\hbox{if $x_0<0$ or $x_0\geq L$}
\eea
and
\be
P_{-}\light(x)\Big|_{x_0=0}=P_{+}\light(x)\Big|_{x_0=L}=0 \,,
\ee
while in the pure gauge part the spatial plaquettes at $x_0=0$ and $x_0=L$
receive a non-trivial (and coupling dependent) weight, see
Eq.~(A.29) of \Ref{zastat:pap1}. 
In the discretization of the SF as described in \cite{SF:LNWW,impr:pap1}
we choose zero boundary gauge fields throughout, $C=C'=0$, which translates 
into the boundary conditions $U(x,k)|_{x_0=0}=U(x,k)|_{x_0=L}=1$ for the 
lattice gauge field.
Similarly, the light and static quark fields at $x_0=0,L$ are fixed to
appropriate (space dependent) boundary functions; the corresponding
boundary conditions are collected in Eqs.~(3.2), (3.3) and (3.5) of
\Ref{zastat:pap1} and not repeated here.
\subsubsection*{SF correlation functions}
Observables are then defined as usual through a path integral involving
the total action $S$.
In this work we focus on SF correlation functions that are constructed from 
the $\Oa$ improved static-light axial current
\bea
\Astatimpr(x) 
& = &
\Astat(x)+a\castat\delta\Astat(x) \,,\\
\Astat(x)
& = &
\lightb(x)\gamma_0\gamma_5\heavy(x) \,,\\
\delta\Astat(x)
& = &
\lightb(x)\gamma_j\gamma_5{1\over 2}
\left(\lnab{j}+\lnabstar{j}\right)\heavy(x) \,.
\eea
Unless it is indicated differently, 
the improvement coefficient $\castat$ is set to its one-loop
perturbative value,
\be
\castat=-{1 \over 4\pi}\,g_0^2 \,, 
\label{ca_stat}
\ee
computed in \Refs{stat:ca,stat:ca_jp}.
On the lattice, in terms of the boundary quark fields 
$\zeta,\ldots,\zetabarprime$, the correlation functions of these fields, 
as well as the various types of pseudoscalar correlators from one boundary 
to the other that are needed in addition, read explicitly:
\bea
\fastat(x_0)
& = &
-a^6\sum_{\vecy,\vecz}{1\over 2}\mvl{
\Astat(x)\,\zetahb(\vecy)\gamma_5\zetal(\vecz)} \,,
\label{fastat}\\
\fdeltaastat(x_0)
& = &
-a^6\sum_{\vecy,\vecz}{1\over 2}\mvl{
\delta\Astat(x)\,\zetahb(\vecy)\gamma_5\zetal(\vecz)} \,,
\label{fdastat}\\
\fone 
& = & 
-{{a^{12}}\over{L^6}}\sum_{\vecu,\vecv,\vecy,\vecz}{1\over 2}
\mvl{\zetaibprime(\vecu)\gamma_5\zetajprime(\vecv)\,
\zetabar_j(\vecy)\gamma_5\zeta_i(\vecz)} \,,
\label{f1}\\
\fonestat
& = &
-{{a^{12}}\over{L^6}}\sum_{\vecu,\vecv,\vecy,\vecz}{1\over 2}
\mvl{\zetalbprime(\vecu)\gamma_5\zetahprime(\vecv)\,
\zetahb(\vecy)\gamma_5\zetal(\vecz)} \,,
\label{f1hl}\\
\fonehh(x_3) 
& = & 
-{{a^{8}}\over{L^2}}\sum_{x_1,x_2,\vecy,\vecz}{1\over 2}\mvl{
\zetahbprime(\vecx)\gamma_5\zetahprime({\bf 0})\,
\zetahb(\vecy)\gamma_5\zetah(\vecz)} \,.
\label{f1hh}
\eea
Moreover we introduce the two ratios
\be
X_{\rm I}(g_0,L/a)=
{{\fastat(L/2)+a\castat\fdeltaastat(L/2)}
\over{\sqrt{\fonestat}}} \,,
\label{XIdef_old}
\ee
\be
\Xnew_{\rm I}(g_0,L/a)=
{{\fastat(L/2)+a\castat\fdeltaastat(L/2)}
\over{\left[f_1\,\fonehh(L/2)\right]^{1/4}}} \,,
\label{XIdef}
\ee
which are constructed such that the (unknown) wave function renormalization 
factors of the boundary quark fields as well as the (linearly divergent) 
mass counterterm $\delta m$ cancel out and only the static current remains 
subject to renormalization.
\subsubsection*{Renormalization}
In \Ref{zastat:pap1} the renormalization constant $\zastat\equiv\zastatSF$ 
entering the $\Oa$ improved static axial current renormalized in the SF 
scheme,
\be
\Aren= 
\zastat(1+\bastat a\mq)\Astatimpr\,,
\ee
was defined in terms of the ratio \Eq{XIdef_old} by imposing the 
renormalization condition (with $m_0$, $\mq$ and $\mc$ as defined in 
\cite{zastat:pap1})
\be
\zastatold(g_0,L/a)\,X_{\rm I}(g_0,L/a)=X_{\rm I}(0,L/a)  
\quad{\rm at}\quad m_0=\mc \,. 
\label{SFdef_old}
\ee
Thus, $\zastatold$ naturally runs with
the scale $\mu=1/L$.
In the present context it will be referred to as the `old' scheme, whereas
the so-called `new' scheme based on \Eq{XIdef} is specified by
\be
\zastat(g_0,L/a)\,\Xnew_{\rm I}(g_0,L/a)=\Xnew_{\rm I}(0,L/a) \,,
\quad m_0=\mc \,, \quad L=1/\mu \,.
\label{SFdef}
\ee
For $\theta=0.5$, which is chosen in our simulations, the relevant values
of the tree-level normalization constant $\Xnew_{\rm I}(0,L/a)$ 
(or $\Xi_{\rm I}^{(0)}(a/L)$ in the notation of \APP{App_pert} 
summarizing the perturbative calculations) are collected 
in \Tab{tab:Xtree_res}.
As an aside we remark that $\Xnew_{\rm I}(0,L/a)=X_{\rm I}(0,L/a)$ holds.
%
%
\begin{table}[htb]
\centering
\begin{tabular}{ccccc}
\hline \\[-2.0ex]
  $L/a$ &&& $\Xnew_{\rm I}(0,L/a)$ \\[1.0ex]
\hline \\[-2.0ex]
   6 &&& $-1.5964837518021$        \\
   8 &&& $-1.5996643156321$        \\
  10 &&& $-1.6011462370857$        \\
  12 &&& $-1.6019540566018$        \\
  16 &&& $-1.6027594410020$        \\
  20 &&& $-1.6031330222299$        \\
  24 &&& $-1.6033361949926$        \\
  32 &&& $-1.6035384024722$        \\[1.0ex]
\hline \\[-2.0ex]
\end{tabular}
\caption{{\sl 
The ratio $\Xnew_{\rm I}$ for $\theta=0.5$ at tree-level.
}}\label{tab:Xtree_res}
\end{table}

The critical quark mass is always understood to be defined from the 
non-perturbatively $\Oa$ improved PCAC mass in the light quark sector as 
in \Ref{msbar:pap1} (i.e.~for $\theta=0$ and $T=L$, evaluating the
associated combination of correlation functions at $x_0=T/2$).
\subsubsection*{Lattice step scaling function}
The lattice step scaling function of the static axial current is defined 
through
\be
\SigmaAstat(u,a/L)=
{{\zastat(g_0,2L/a)}\over{\zastat(g_0,L/a)}} 
\quad{\rm at}\quad \gbar^2(L)=u\,,m_0=\mc \,.
\label{SSF_lat}
\ee
The additional condition $m_0=\mc$ from above, referring to lattice size
$L/a$, defines the critical hopping parameter value, $\kappa=\hopc$.
Moreover, enforcing $\gbar^2(L)$ to take some prescribed value $u$ fixes 
the bare coupling value $g_0^2=6/\beta$ to be used for given $L/a$.
In this way $\SigmaAstat$ becomes a function of the renormalized coupling
$u$, up to cutoff effects, and approaches its continuum limit as 
$a/L\rightarrow 0$ for fixed $u$.
\subsection{Simulation details and results}
\label{App_latSSF_res}
As emphasized before, our quenched lattice simulation and the data analysis 
are analogous to \Ref{msbar:pap1}, except that the boundary coefficient 
$\ct$ is set to its two-loop perturbative value \cite{impr:ct_2loop}:
\be
\ct^{{\rm 2-loop}}=1-0.089\,g_0^2-0.030\,g_0^4 \,.
\label{ct_2loop}
\ee
The boundary $\Oa$ improvement terms involving quark fields have to be
multiplied with a coefficient $\cttil$, which is known to one-loop 
from \cite{impr:pap2}, viz.
\be
\cttil^{\,{\rm 1-loop}}=1-0.018\,g_0^2 \,.
\label{cttil_1loop}
\ee
Of course, owing to a priori unknown precision to which perturbation
theory approximates these coefficients, linear lattice spacing errors are 
not suppressed completely, and we will come back to this issue later.
As for the other contributing $\Oa$ improvement coefficients, we used the
non-perturbative values for $\csw$ and $\ca$ of \cite{impr:pap3} for the 
relativistic fermions and the one-loop estimate (\ref{ca_stat}) 
for $\castat$ in 
the static-light axial current.

The renormalization constants $\zastat(g_0,L/a)$ and $\zastat(g_0,2L/a)$ 
in \Eq{SSF_lat} have been evaluated from the correlation functions 
in \Eqs{fastat} -- (\ref{f1hh}), which were computed in a numerical 
simulation with $\theta=0.5$. (The latter parameter specifies the boundary 
conditions of the quark fields, see e.g.~\cite{pert:1loop}.)
These simulations were performed on the APE-100 parallel computers with
128 to 512 nodes, employing for the updating of the gauge fields the same 
hybrid-overrelaxation algorithm as in \cite{alpha:SU3,msbar:pap1} with
two overrelaxation sweeps per heatbath sweep within a full iteration.
This mix of updating was found to be close to optimal in \cite{thesis:berndg}.
Since the computation of SF correlators has already been detailed in
\Refs{impr:pap3} and (Appendix~A.2.2 of) \cite{msbar:pap1}, we just mention
that we differ from them only by using the implementation 
\cite{ssor:impr_ALPHA} of the SSOR-preconditioned BiCGStab
inverter \cite{ssor:wils_wupp} to solve the lattice Dirac equation.

The computation of $\fonehh(d)$, where $d=|x_3|$ and $a \le d \le L/2$, 
amounts to evaluate \Eq{e_fonehh_app}. 
In order to improve the statistical precision of $\fonehh$, the links 
building up the observable are evaluated by a 10--hit multi-hit 
procedure \cite{PPR}, where each hit consists of a Cabibbo-Marinari 
heatbath update in three SU(2)--subgroups of SU(3). Translation invariance
is fully exploited. 

In order to keep autocorrelations small, the measurements of the 
correlation functions were always separated by $L/(2a)$ update iterations 
(and, respectively, five iterations in the case of $\fonehh$). 
Statistical errors stem from a standard jackknife analysis, where we have
checked explicitly for the statistical independence of the data by
averaging them into bins of a few consecutive measurements beforehand.
The total number of measurements itself was such that the statistical 
error of $\SigmaAstat$ was dominated by the uncertainty in 
$\zastat(g_0,2L/a)$.
In general, the uncertainties in the coupling and in the value of $\hopc$
would have to be propagated into the error of $\SigmaAstat$ as well.
But as the former can be estimated 
to be much smaller than the statistical error of
$\SigmaAstat$ and $\SigmaAstat$ is found to depend rather weakly on the 
bare (light) quark mass, we neglected both contributions in the final 
error estimate. 

In \Tab{tab:zp_res} -- \Tab{tab:zastat2_res} we list our results 
on the
step scaling functions of the static axial current\footnote{
Here we do not tabulate the results on the static-static boundary correlator
$\fonehh$ separately, but the numbers can be obtained from the authors 
upon request.
} 
and 
--- since they are available from our computations as well ---
of the pseudoscalar density defined as in
\cite{msbar:pap1}.
The values of 
$\beta$ and the critical hopping parameter $\kappa=\hopc$ to be 
simulated were taken over from \Ref{msbar:pap1} without changes, which
means to stay with $\ct$ and $\cttil$ to one-loop accuracy in realizing the 
conditions $\gbar^2(L)=u$ and $m_0=\mc$.
Note once more, however, that here, in contrast to \cite{msbar:pap1}, for 
the corresponding renormalization constants themselves --- particularly 
when comparing the results for $\zp$ and $\sigplatt$ quoted in that 
previous work with those of the present one --- the two-loop formula for 
$\ct$, \Eq{ct_2loop}, has been used.
\clearpage
%
%
\begin{table}[htb]
\centering
\begin{tabular}{lcrcrlll}
\hline
  $\gbar^2(L)$ && $\beta$ & $\kappa$ & $L/a$ & $\zp(g_0,L/a)$ 
& $\zp(g_0,2L/a)$ & $\Sigmap(u,a/L)$ \\
\hline
\input table_SigZP.tex
\hline \\[-2.0ex]
\end{tabular}
\caption{{\sl
Results for the step scaling function $\Sigmap$.
}}\label{tab:zp_res}
\end{table}
%
%
\begin{table}[htb]
\centering
\begin{tabular}{lcrcrlll}
\hline
  $\gbar^2(L)$ && $\beta$ & $\kappa$ & $L/a$ & $\zastat(g_0,L/a)$ 
& $\zastat(g_0,2L/a)$ & $\SigmaAstat(u,a/L)$ \\
\hline
\input table_SigZAstat1.tex
\hline \\[-2.0ex]
\end{tabular}
\caption{{\sl 
Results for the step scaling function $\SigmaAstat$ (in the
`new' scheme).
}}\label{tab:zastat1_res}
\end{table}
%
%
\begin{table}[htb]
\centering
\begin{tabular}{lcrcrlll}
\hline
  $\gbar^2(L)$ && $\beta$ & $\kappa$ & $L/a$ & $\zastatold(g_0,L/a)$ 
& $\zastatold(g_0,2L/a)$ & $\SigmaAstatold(u,a/L)$ \\
\hline
\input table_SigZAstat2.tex
\hline \\[-2.0ex]
\end{tabular}
\caption{{\sl 
Results for the step scaling function $\SigmaAstatold$ (in the
`old' scheme).
}}\label{tab:zastat2_res}
\end{table}
\clearpage
\subsection{Continuum extrapolation of $\SigmaAstat$}
\label{App_latSSF_extrap}
For fixed coupling $u$ the step scaling function defined 
in \Eq{SSF_lat} has a continuum limit, $\sigmaAstat(u)$.
Neglecting for the moment the uncertainties on the correct values
of $\ct$, $\cttil$ and $\castat$, we expect the leading-order cutoff 
effects to be quadratic in the lattice spacing,
\be
\SigmaAstat(u,a/L)=\sigmaAstat(u)+\rmO(a^2/L^2) \,,
\ee
since $\Oa$ improvement is employed.
Based on this ansatz, \Fig{fig:Sigma_zastat1} in \Sect{Sec_NPrun} 
illustrates the continuum extrapolation of $\SigmaAstat$ for a 
representative subset of our available coupling values $u=\gbar^2(L)$. 
The coarsest lattices (with $L/a=6$) have been omitted from the fits as a 
safeguard against higher order cutoff effects. For the remaining 
$a/L\leq 1/8$, the one-loop cutoff effects are quite small, 
see \Fig{f_delta}.

Although these extrapolations are entirely compatible with an approach to 
the continuum limit quadratic in $a/L$, we also have investigated 
extrapolations linear in $a/L$.
These as well yield reasonable fits with consistent results and even
comparable total $\chi^2/{\rm dof}$ (when summing up the $\chi^2$-s 
belonging to the individual fits at the ten $u$--values) so that 
the form of the lattice spacing 
dependence can not be decided on the basis of the data.
Therefore, we have studied the influence of the imperfect (i.e.~only 
perturbative) knowledge of some of the improvement coefficients in more 
detail. 

Since the usage of the two-loop approximation (\ref{ct_2loop}) for $\ct$ 
in the calculation of the correlation functions (and thereby also in the 
step scaling function $\SigmaAstat$) should cancel the main contributions 
from the related boundary terms, we only address its effect originating
from the fixing of the renormalized coupling, the values of which were 
taken over from \Ref{msbar:pap1} with $\ct$ still set to one-loop.
Changing $\ct$ from one- to two-loop also in this step then requires to 
adjust the bare coupling and the value of the critical quark mass
accordingly before the simulations for $\SigmaAstat$ can be performed.
We have done this analysis for the largest fixed coupling, $u=3.48$, where 
the uncertainty in $\ct$ is largest and thus its effect most pronounced. 
At $L/a=6$ the resulting change in $\SigmaAstat$ turns out to lie clearly 
inside the statistical errors, and this effect will even get smaller for 
decreasing $a/L$.
On the other hand, if we just compute $\SigmaAstat$ with the one-loop 
value of $\ct$ as in in the computation of the coupling, we found the 
results, now for $u=2.77$ and $L/a=6,8$, to be indistinguishable within
errors, too.
We conclude that any small uncertainty present in $\ct$ beyond the 
available two-loop estimate is numerically unimportant for the cutoff 
dependence of $\SigmaAstat$.

Regarding the $\Oa$ improvement coefficient $\cttil$, we followed 
the same line as in \cite{msbar:pap1}
and assessed its influence on our results by artificially replacing the
one-loop coefficient in the expression (\ref{cttil_1loop}) by ten times
its value.
I.e.~we set $\cttil$ to $\cttil\kern1pt'=1-0.180\,g_0^2\,$
in some additional simulations at $u=3.48$, and the outcome is that the 
corresponding estimates on $\SigmaAstat$ for $L/a=6$ still differ by 
around $1.5\%$, while for $L/a=8$ they already agree within their 
statistical errors.
As this difference drops further for growing $L/a$ and/or smaller 
couplings, a possible imperfection of $\cttil$ does not affect the results
on $\SigmaAstat$ either.  

Finally, we also checked for the influence of the $\Oa$ improvement
coefficient in the static-light axial current, $\castat$, by analyzing our
data with $\castat=0$ instead of the one-loop value (\ref{ca_stat}).
Whereas the related change in $\zastat$ is of the order of a few percent
and hence still substantial, it largely cancels in the ratio
of \Eq{SSF_lat} so that this effect is no more significant for 
$\SigmaAstat$ given its statistical errors.

All in all these findings demonstrate that at the level of our precision 
linear $a$--effects in the data on $\SigmaAstat$ are negligible, and 
extrapolations using $(a/L)^2$--terms as the dominant scaling violation are 
justified indeed.

%% file: table_SigZP.tex
 1.0989 &&
 9.5030 & 0.131514 &  6 & $0.8190(10)$ & $0.7846(9 )$ & $0.9581(16)$
\\
        &&
 9.7500 & 0.131312 &  8 & $0.8107(9 )$ & $0.7786(12)$ & $0.9604(18)$
\\
        &&
10.0577 & 0.131079 & 12 & $0.8012(6 )$ & $0.7702(11)$ & $0.9613(16)$
\\
        &&
10.3419 & 0.130876 & 16 & $0.7956(8 )$ & $0.7625(15)$ & $0.9584(21)$
\\
\hline
 1.3293 &&
 8.6129 & 0.132380 &  6 & $0.7930(10)$ & $0.7493(10)$ & $0.9449(17)$
\\
        &&
 8.8500 & 0.132140 &  8 & $0.7827(11)$ & $0.7402(11)$ & $0.9456(19)$
\\
        &&
 9.1859 & 0.131814 & 12 & $0.7737(7 )$ & $0.7353(12)$ & $0.9503(17)$
\\
        &&
 9.4381 & 0.131589 & 16 & $0.7666(11)$ & $0.7274(18)$ & $0.9488(27)$
\\
\hline
 1.4300 &&
 8.5598 & 0.132453 &  8 & $0.7702(10)$ & $0.7273(13)$ & $0.9443(21)$
\\
        &&
 8.9003 & 0.132095 & 12 & $0.7610(6 )$ & $0.7223(14)$ & $0.9491(21)$
\\
        &&
 9.1415 & 0.131855 & 16 & $0.7555(7 )$ & $0.7123(20)$ & $0.9428(28)$
\\
\hline
 1.5553 &&
 7.9993 & 0.133118 &  6 & $0.7666(7 )$ & $0.7165(16)$ & $0.9346(23)$
\\
        &&
 8.2500 & 0.132821 &  8 & $0.7590(8 )$ & $0.7134(13)$ & $0.9399(20)$
\\
        &&
 8.5985 & 0.132427 & 12 & $0.7473(10)$ & $0.7035(13)$ & $0.9414(21)$
\\
        &&
 8.8323 & 0.132169 & 16 & $0.7421(9 )$ & $0.6976(19)$ & $0.9401(29)$
\\
\hline
 1.6950 &&
 7.9741 & 0.133179 &  8 & $0.7442(11)$ & $0.6939(15)$ & $0.9325(24)$
\\
        &&
 8.3218 & 0.132756 & 12 & $0.7341(7 )$ & $0.6862(15)$ & $0.9348(22)$
\\
        &&
 8.5479 & 0.132485 & 16 & $0.7277(13)$ & $0.6805(18)$ & $0.9352(30)$
\\
\hline
 1.8811 &&
 7.4082 & 0.133961 &  6 & $0.7348(9 )$ & $0.6764(6 )$ & $0.9205(14)$
\\
        &&
 7.6547 & 0.133632 &  8 & $0.7258(7 )$ & $0.6691(15)$ & $0.9219(23)$
\\
        &&
 7.9993 & 0.133159 & 12 & $0.7173(5 )$ & $0.6632(8 )$ & $0.9245(12)$
\\
        &&
 8.2415 & 0.132847 & 16 & $0.7117(13)$ & $0.6604(20)$ & $0.9279(33)$
\\
\hline
 2.1000 &&
 7.3632 & 0.134088 &  8 & $0.7088(13)$ & $0.6433(16)$ & $0.9076(28)$
\\
        &&
 7.6985 & 0.133599 & 12 & $0.6971(8 )$ & $0.6385(24)$ & $0.9160(36)$
\\
        &&
 7.9560 & 0.133229 & 16 & $0.6919(12)$ & $0.6303(17)$ & $0.9110(29)$
\\
\hline
 2.4484 &&
 6.7807 & 0.134994 &  6 & $0.6845(10)$ & $0.6110(12)$ & $0.8925(21)$
\\
        &&
 7.0197 & 0.134639 &  8 & $0.6784(8 )$ & $0.6061(19)$ & $0.8933(30)$
\\
        &&
 7.2025 & 0.134380 & 10 & $0.6733(8 )$ & $0.6021(12)$ & $0.8943(21)$
\\
        &&
 7.3551 & 0.134141 & 12 & $0.6722(11)$ & $0.6012(12)$ & $0.8944(24)$
\\
        &&
 7.6101 & 0.133729 & 16 & $0.6661(5 )$ & $0.5962(10)$ & $0.8950(17)$
\\
\hline
 2.7700 &&
 6.5512 & 0.135327 &  6 & $0.6619(10)$ & $0.5758(20)$ & $0.8699(33)$
\\
        &&
 6.7860 & 0.135056 &  8 & $0.6541(13)$ & $0.5751(17)$ & $0.8792(31)$
\\
        &&
 6.9720 & 0.134770 & 10 & $0.6505(8 )$ & $0.5717(17)$ & $0.8788(28)$
\\
        &&
 7.1190 & 0.134513 & 12 & $0.6482(9 )$ & $0.5705(10)$ & $0.8802(19)$
\\
        &&
 7.3686 & 0.134114 & 16 & $0.6442(16)$ & $0.5668(16)$ & $0.8798(33)$
\\
\hline
 3.4800 &&
 6.2204 & 0.135470 &  6 & $0.6173(8 )$ & $0.5067(11)$ & $0.8208(21)$
\\
        &&
 6.4527 & 0.135543 &  8 & $0.6133(8 )$ & $0.5101(21)$ & $0.8316(35)$
\\
        &&
 6.6350 & 0.135340 & 10 & $0.6112(11)$ & $0.5078(19)$ & $0.8307(35)$
\\
        &&
 6.7750 & 0.135121 & 12 & $0.6076(7 )$ & $0.5061(14)$ & $0.8329(24)$
\\
        &&
 7.0203 & 0.134707 & 16 & $0.6063(7 )$ & $0.5097(11)$ & $0.8406(21)$
\\

%% file: table_SigZAstat1.tex
 1.0989 &&
 9.5030 & 0.131514 &  6 & $0.8926(7 )$ & $0.8698(8 )$ & $0.9745(12)$
\\
        &&
 9.7500 & 0.131312 &  8 & $0.8860(7 )$ & $0.8668(11)$ & $0.9782(14)$
\\
        &&
10.0577 & 0.131079 & 12 & $0.8800(6 )$ & $0.8630(11)$ & $0.9806(14)$
\\
        &&
10.3419 & 0.130876 & 16 & $0.8786(7 )$ & $0.8586(16)$ & $0.9773(20)$
\\
\hline
 1.3293 &&
 8.6129 & 0.132380 &  6 & $0.8733(8 )$ & $0.8458(9 )$ & $0.9686(13)$
\\
        &&
 8.8500 & 0.132140 &  8 & $0.8677(9 )$ & $0.8418(13)$ & $0.9701(18)$
\\
        &&
 9.1859 & 0.131814 & 12 & $0.8635(7 )$ & $0.8401(13)$ & $0.9729(17)$
\\
        &&
 9.4381 & 0.131589 & 16 & $0.8593(9 )$ & $0.8361(19)$ & $0.9731(25)$
\\
\hline
 1.4300 &&
 8.5598 & 0.132453 &  8 & $0.8593(7 )$ & $0.8328(13)$ & $0.9692(17)$
\\
        &&
 8.9003 & 0.132095 & 12 & $0.8545(7 )$ & $0.8314(14)$ & $0.9731(18)$
\\
        &&
 9.1415 & 0.131855 & 16 & $0.8516(6 )$ & $0.8238(23)$ & $0.9674(28)$
\\
\hline
 1.5553 &&
 7.9993 & 0.133118 &  6 & $0.8572(6 )$ & $0.8246(13)$ & $0.9619(16)$
\\
        &&
 8.2500 & 0.132821 &  8 & $0.8517(6 )$ & $0.8248(13)$ & $0.9684(17)$
\\
        &&
 8.5985 & 0.132427 & 12 & $0.8459(8 )$ & $0.8190(14)$ & $0.9683(20)$
\\
        &&
 8.8323 & 0.132169 & 16 & $0.8425(9 )$ & $0.8141(20)$ & $0.9662(26)$
\\
\hline
 1.6950 &&
 7.9741 & 0.133179 &  8 & $0.8414(9 )$ & $0.8069(18)$ & $0.9590(24)$
\\
        &&
 8.3218 & 0.132756 & 12 & $0.8359(8 )$ & $0.8074(13)$ & $0.9659(19)$
\\
        &&
 8.5479 & 0.132485 & 16 & $0.8329(13)$ & $0.8081(19)$ & $0.9703(28)$
\\
\hline
 1.8811 &&
 7.4082 & 0.133961 &  6 & $0.8362(7 )$ & $0.7939(7 )$ & $0.9495(11)$
\\
        &&
 7.6547 & 0.133632 &  8 & $0.8290(6 )$ & $0.7903(17)$ & $0.9533(21)$
\\
        &&
 7.9993 & 0.133159 & 12 & $0.8247(7 )$ & $0.7907(12)$ & $0.9588(16)$
\\
        &&
 8.2415 & 0.132847 & 16 & $0.8221(13)$ & $0.7898(23)$ & $0.9607(32)$
\\
\hline
 2.1000 &&
 7.3632 & 0.134088 &  8 & $0.8193(10)$ & $0.7732(20)$ & $0.9436(26)$
\\
        &&
 7.6985 & 0.133599 & 12 & $0.8117(9 )$ & $0.7756(25)$ & $0.9555(32)$
\\
        &&
 7.9560 & 0.133229 & 16 & $0.8081(12)$ & $0.7704(21)$ & $0.9533(29)$
\\
\hline
 2.4484 &&
 6.7807 & 0.134994 &  6 & $0.8035(8 )$ & $0.7420(12)$ & $0.9235(18)$
\\
        &&
 7.0197 & 0.134639 &  8 & $0.7945(7 )$ & $0.7444(19)$ & $0.9370(26)$
\\
        &&
 7.2025 & 0.134380 & 10 & $0.7936(9 )$ & $0.7431(17)$ & $0.9363(24)$
\\
        &&
 7.3551 & 0.134141 & 12 & $0.7930(10)$ & $0.7465(17)$ & $0.9414(24)$
\\
        &&
 7.6101 & 0.133729 & 16 & $0.7907(8 )$ & $0.7444(16)$ & $0.9415(22)$
\\
\hline
 2.7700 &&
 6.5512 & 0.135327 &  6 & $0.7886(9 )$ & $0.7133(19)$ & $0.9045(26)$
\\
        &&
 6.7860 & 0.135056 &  8 & $0.7791(12)$ & $0.7187(24)$ & $0.9225(35)$
\\
        &&
 6.9720 & 0.134770 & 10 & $0.7786(9 )$ & $0.7223(23)$ & $0.9276(31)$
\\
        &&
 7.1190 & 0.134513 & 12 & $0.7740(11)$ & $0.7220(17)$ & $0.9329(25)$
\\
        &&
 7.3686 & 0.134114 & 16 & $0.7755(16)$ & $0.7281(30)$ & $0.9388(43)$
\\
\hline
 3.4800 &&
 6.2204 & 0.135470 &  6 & $0.7587(10)$ & $0.6562(39)$ & $0.8649(53)$
\\
        &&
 6.4527 & 0.135543 &  8 & $0.7496(11)$ & $0.6624(29)$ & $0.8837(41)$
\\
        &&
 6.6350 & 0.135340 & 10 & $0.7477(11)$ & $0.6658(31)$ & $0.8906(43)$
\\
        &&
 6.7750 & 0.135121 & 12 & $0.7451(11)$ & $0.6696(22)$ & $0.8988(32)$
\\
        &&
 7.0203 & 0.134707 & 16 & $0.7470(10)$ & $0.6792(24)$ & $0.9092(34)$
\\

%% file: table_SigZAstat2.tex
 1.0989 &&
 9.5030 & 0.131514 &  6 & $0.8903(8 )$ & $0.8676(8 )$ & $0.9745(12)$
\\
        &&
 9.7500 & 0.131312 &  8 & $0.8846(7 )$ & $0.8635(11)$ & $0.9762(14)$
\\
        &&
10.0577 & 0.131079 & 12 & $0.8791(5 )$ & $0.8628(14)$ & $0.9814(17)$
\\
        &&
10.3419 & 0.130876 & 16 & $0.8773(9 )$ & $0.8601(21)$ & $0.9804(26)$
\\
\hline
 1.3293 &&
 8.6129 & 0.132380 &  6 & $0.8721(9 )$ & $0.8430(10)$ & $0.9666(16)$
\\
        &&
 8.8500 & 0.132140 &  8 & $0.8664(10)$ & $0.8428(14)$ & $0.9728(19)$
\\
        &&
 9.1859 & 0.131814 & 12 & $0.8616(6 )$ & $0.8351(19)$ & $0.9693(23)$
\\
        &&
 9.4381 & 0.131589 & 16 & $0.8580(10)$ & $0.8370(48)$ & $0.9755(57)$
\\
\hline
 1.4300 &&
 8.5598 & 0.132453 &  8 & $0.8577(8 )$ & $0.8325(14)$ & $0.9706(19)$
\\
        &&
 8.9003 & 0.132095 & 12 & $0.8535(6 )$ & $0.8332(21)$ & $0.9761(26)$
\\
        &&
 9.1415 & 0.131855 & 16 & $0.8488(6 )$ & $0.8172(55)$ & $0.9628(65)$
\\
\hline
 1.5553 &&
 7.9993 & 0.133118 &  6 & $0.8552(7 )$ & $0.8243(19)$ & $0.9639(23)$
\\
        &&
 8.2500 & 0.132821 &  8 & $0.8497(6 )$ & $0.8224(17)$ & $0.9678(22)$
\\
        &&
 8.5985 & 0.132427 & 12 & $0.8441(8 )$ & $0.8198(26)$ & $0.9712(32)$
\\
        &&
 8.8323 & 0.132169 & 16 & $0.8410(12)$ & $0.8203(57)$ & $0.9754(69)$
\\
\hline
 1.6950 &&
 7.9741 & 0.133179 &  8 & $0.8411(11)$ & $0.8079(20)$ & $0.9606(26)$
\\
        &&
 8.3218 & 0.132756 & 12 & $0.8340(8 )$ & $0.8154(32)$ & $0.9777(39)$
\\
        &&
 8.5479 & 0.132485 & 16 & $0.8328(16)$ & $0.8076(87)$ & $0.970 (11)$
\\
\hline
 1.8811 &&
 7.4082 & 0.133961 &  6 & $0.8336(7 )$ & $0.7904(7 )$ & $0.9482(12)$
\\
        &&
 7.6547 & 0.133632 &  8 & $0.8269(7 )$ & $0.7921(21)$ & $0.9578(27)$
\\
        &&
 7.9993 & 0.133159 & 12 & $0.8232(5 )$ & $0.7863(20)$ & $0.9551(25)$
\\
        &&
 8.2415 & 0.132847 & 16 & $0.8210(18)$ & $0.798 (12)$ & $0.972 (15)$
\\
\hline
 2.1000 &&
 7.3632 & 0.134088 &  8 & $0.8172(13)$ & $0.7778(24)$ & $0.9518(33)$
\\
        &&
 7.6985 & 0.133599 & 12 & $0.8097(8 )$ & $0.7757(67)$ & $0.9579(83)$
\\
        &&
 7.9560 & 0.133229 & 16 & $0.8091(21)$ & $0.786 (12)$ & $0.971 (14)$
\\
\hline
 2.4484 &&
 6.7807 & 0.134994 &  6 & $0.8016(9 )$ & $0.7416(17)$ & $0.9252(23)$
\\
        &&
 7.0197 & 0.134639 &  8 & $0.7941(9 )$ & $0.7399(37)$ & $0.9317(48)$
\\
        &&
 7.2025 & 0.134380 & 10 & $0.7932(9 )$ & $0.7422(37)$ & $0.9357(48)$
\\
        &&
 7.3551 & 0.134141 & 12 & $0.7901(13)$ & $0.7382(54)$ & $0.9344(70)$
\\
        &&
 7.6101 & 0.133729 & 16 & $0.7870(9 )$ & $0.756 (13)$ & $0.961 (17)$
\\
\hline
 2.7700 &&
 6.5512 & 0.135327 &  6 & $0.7863(10)$ & $0.7132(34)$ & $0.9070(45)$
\\
        &&
 6.7860 & 0.135056 &  8 & $0.7804(15)$ & $0.7169(45)$ & $0.9185(60)$
\\
        &&
 6.9720 & 0.134770 & 10 & $0.7759(10)$ & $0.7121(55)$ & $0.9177(72)$
\\
        &&
 7.1190 & 0.134513 & 12 & $0.7739(12)$ & $0.7152(72)$ & $0.9241(95)$
\\
        &&
 7.3686 & 0.134114 & 16 & $0.7732(30)$ & $0.681 (34)$ & $0.880 (43)$
\\
\hline
 3.4800 &&
 6.2204 & 0.135470 &  6 & $0.7573(9 )$ & $0.6558(24)$ & $0.8659(34)$
\\
        &&
 6.4527 & 0.135543 &  8 & $0.7501(9 )$ & $0.6560(77)$ & $0.874 (10)$
\\
        &&
 6.6350 & 0.135340 & 10 & $0.7476(15)$ & $0.661 (11)$ & $0.885 (15)$
\\
        &&
 6.7750 & 0.135121 & 12 & $0.7430(10)$ & $0.659 (16)$ & $0.886 (21)$
\\
        &&
 7.0203 & 0.134707 & 16 & $0.7474(13)$ & $0.642 (46)$ & $0.859 (61)$
\\

%% file: apdxb.tex
\section{Perturbation theory}
\label{App_pert}
This appendix provides a few details on the perturbative computations, 
which were required to obtain the one-loop expression for $h(d/L,u)$, 
\Eq{e_h}, the two-loop anomalous dimension and the one-loop
estimates of the discretization errors of the step scaling function
$\SigmaAstat$.
Note that here we restrict ourselves to the case of the modified 
(or `new') scheme introduced via \Eq{rencond} in this paper, because 
the perturbation theory of the original scheme defined 
through \Eq{rencond_old} has been extensively discussed already 
in \Ref{zastat:pap1} where also more details on the different steps 
involved can be found.

The correlation functions $\fastat,\;\fdeltaastat$ and $\fone$ are 
expanded in powers of
the coupling $g_0^2$ as explained in~\cite{zastat:pap1} 
and~\cite{impr:pap5}, and the analogous expansion of
$\fonehh$ is explained below.
\subsection{The correlation function $\fonehh$}
\label{App_pert_f1hh}
After integrating out the static quark fields,
the correlation function $\fonehh$ can be written as
\begin{eqnarray}
\fonehh(x_3) 
& = & 
{{a^2}\over{L^2}}\sum_{x_1,x_2}\bigg\langle\tr\Big\{
U(x,0)U(x+a\hat{0},0)\cdots U(x+(L-a)\hat{0},0) \nonumber\\
&   &
\times\,U((L-a)\hat{0},0)^{-1}U((L-2a)\hat{0},0)^{-1}\cdots  
  U(0,0)^{-1}\Big\}\Big|_{\,x_0=0}\bigg\rangle \,,
\label{e_fonehh_app}
\end{eqnarray}
where the trace is taken over colour indices only.

Writing $U(x,\mu)=\exp\left\{g_0aq_{\mu}(x)\right\}$,
with the gluon field $q_{\mu}(x)=q_{\mu}^a(x)T^a$,
where $T^a$ are the antihermitian generators of the gauge group, 
the function $\fonehh$ can be expanded in the bare coupling,
\begin{equation}
\fonehh(d)=3 + g_0^2\fonehhoneloop(d)+{\rm O}(g_0^4) \,,
\quad d=|x_3| \,.
\end{equation}

\begin{figure}[htb]
\begin{center}
\begin{minipage}[b]{0.3\linewidth}
\centering\epsfig{figure=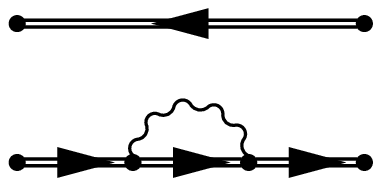,width=\linewidth}
\end{minipage}
\begin{minipage}[b]{0.3\linewidth}
\centering\epsfig{figure=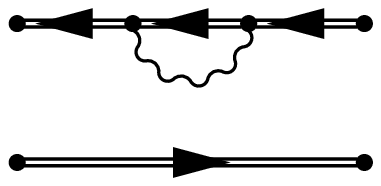,width=\linewidth}
\end{minipage}\\
\vspace{4mm}
\begin{minipage}[b]{0.3\linewidth}
\centering\epsfig{figure=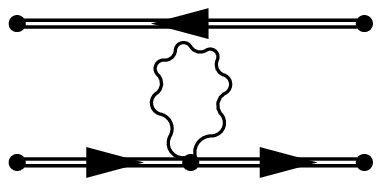,width=\linewidth}
\end{minipage}
\begin{minipage}[b]{0.3\linewidth}
\centering\epsfig{figure=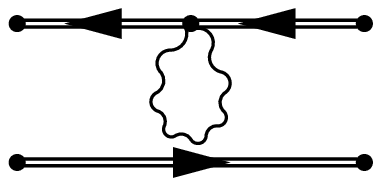,width=\linewidth}
\end{minipage}
\begin{minipage}[b]{0.3\linewidth}
\centering\epsfig{figure=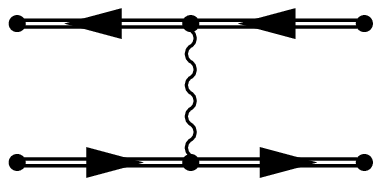,width=\linewidth}
\end{minipage}
\end{center}
\vspace{-1.25cm}
\caption{{\sl
One-loop diagrams contributing to $\fonehh$. 
The two dots on the left are at $x_0=0$, the dots on the right are at 
$x_0=L$.
}}\label{f_f1hh_oneloop}
\end{figure}

Determining the one-loop coefficient $\fonehhoneloop$ amounts to 
calculating and summing the diagrams shown in \Fig{f_f1hh_oneloop}
using the gluon propagator $D_{\mu\nu}(x,y)$ (with SF boundary conditions)
given in \Ref{impr:pap2}.
%
%

For comparison with the non-perturbative results shown in \Fig{fig:f1hh_d},
Subsection~\ref{Sec_rscheme_f1hh}, we consider the 
one-loop coefficient $h^{(1)}(d/L)$ of \Eq{e_hpert}, which reads
\begin{equation}
h^{(1)}(d/L)={1\over 3}\left\{\fonehhoneloop(d)
-\fonehhoneloop(L/2)\right\}\,,
\label{f1hh_ratio}
\end{equation}
and which we can obtain analytically. (For the other quantities
considered in this appendix,
the diagrams are calculated numerically.)
Only the last of the diagrams in \Fig{f_f1hh_oneloop} contributes,
and we thus can write
\begin{eqnarray}
h^{(1)}(d/L) & =  & 
{{4a^6}\over{3L^4}}\sum_{x_0,y_0}\sum_{x_1,y_1}\sum_{x_2,y_2}
\left\{D_{00}(x,y)|_{\,x_3=d\,,\,y_3=0}
-D_{00}(x,y)|_{\,x_3=L/2\,,\,y_3=0}\right\} 
\nonumber\\
& = & 
{{4a^2}\over{3L^3}}\sum_{p_3}\left\{\rme^{\,ip_3d}-\rme^{\,ip_3L/2}\right\}
\sum_{x_0,y_0}d_{00}(x_0,y_0;(0,0,p_3)) \,,
\end{eqnarray}
with the momentum-space gluon propagator $d_{00}(x_0,y_0;{\bf p})$ defined 
in~\Ref{impr:pap2}.
The $p_3=0$ term does not contribute to the sum, and using the explicit 
form of $d_{00}(x_0,y_0;{\bf p})$ for ${\bf p}\neq {\bf 0}$, one can show 
that
\begin{equation}
a^2\sum_{x_0,y_0}d_{00}(x_0,y_0;(0,0,p_3))={L\over{{\hat{p}_3}^2}}
\end{equation}
with $\hat{p}_3={2\over a}\sin\left({ap_3\over 2}\right)$.
Thus we see that $h^{(1)}(d/L)$ is just given in terms of the one-dimensional 
scalar propagator
on a periodic lattice with length $L$, which has {\em no lattice artifacts}.
This eventually leads to the result quoted in \Eq{e_honeloop}. The absence
of any lattice spacing dependence is a consequence of the special 
kinematics, namely the summation over $x_1,x_2$, but will of course not
be exact in higher orders of perturbation theory.
\subsection{Anomalous dimensions}
\label{App_pert_gamma}
In order to precisely connect to the RGI current, 
it is important to obtain the 
anomalous dimension of the static-light axial current in the SF scheme at 
two-loop order. 
The anomalous dimension is expanded as
\begin{equation}
\gamma(\gbar)=
-\gbar^2\,\Big\{\,
\gamma_0+\gamma_1^{\rm SF}\gbar^2+{\rm O}(\gbar^4)\,\Big\} \,,
\label{gam_expansion}
\end{equation}
with
$\gamma_0=-{1/(4\pi^2)}$ and
the two-loop anomalous dimension in the SF scheme, $\gamma_1^{\rm SF}$.

With the perturbative expansions for the various correlation functions, 
the ratio $\Xi_{\rm I}$ of \Eq{XIdef} can be written as a series
\begin{equation}
\Xi_{\rm I}(g_0,a/L)=
\Xi_{\rm I}^{(0)}(a/L)+g_0^2\,\Xi_{\rm I}^{(1)}(a/L)+{\rm O}(g_0^4) \,.
\end{equation}
Accordingly, this allows us to expand the SF renormalization constant 
$\zastat=\Xi_{\rm I}^{(0)}/\Xi_{\rm I}$ (see \Eq{SFdef}) as
\begin{equation}
\zastat=\zastattree+g_0^2\zastatoneloop+{\rm O}(g_0^4) \,.
\label{e_zastatexpansion}
\end{equation}

With the one-loop relation between the bare lattice current and the 
renormalized static axial current in the $\msbar$ scheme, 
the anomalous dimension in the $\msbar$ scheme, 
\Eq{e_gammamsbar}~\cite{Ji:1991pr,BroadhGrozin,Gimenez:1992bf}, can be 
converted into the SF scheme. 
The renormalization constant relating the SF scheme and the $\MSbar$ scheme
is obtained from the relation between the SF scheme and the bare lattice 
current,~\Eq{e_zastatexpansion}, the connection of the bare lattice current
and a `matching scheme' \cite{zastat:pap2,BorrPitt} and the 
relation between the latter and the $\msbar$ scheme~\cite{stat:eichhill_za}. 
Here the matching scheme is defined by the requirement that the renormalized 
static-light axial current at scale $\mu=m_{\rm h}$ equals the relativistic 
axial current with a heavy quark mass $m_{\rm h}$ up to terms of 
${\rm O}(1/m_{\rm h})$, and the current in the relativistic theory 
is normalized by current algebra 
(imposing the chiral Ward identities).\footnote{
Since the wording in \Ref{zastat:pap1} is not completely clear on this,
we point out that $A_\msbar^{\rm stat}$ in that reference refers
to what we call the $\msbar$ scheme here as well as in  \Ref{zastat:pap1}.}
Following the steps in \Ref{zastat:pap1}, this analysis finally yields
\begin{eqnarray}
\theta=0.0:\qquad\gamma_1^{\rm SF} & = &
{1\over{(4\pi)^2}}\,\left\{\,0.22(2)-0.0552(13)N_{\rm f}\,\right\}, \\
\theta=0.5:\qquad\gamma_1^{\rm SF} & = &
{1\over{(4\pi)^2}}\,\left\{\,0.10(2)-0.0477(13)N_{\rm f}\,\right\}, \\
\theta=1.0:\qquad\gamma_1^{\rm SF} & = &
{1\over{(4\pi)^2}}\,\left\{\,-0.08(2)-0.0365(13)N_{\rm f}\,\right\}.
\end{eqnarray}
\subsection{Discretization errors}
\label{App_pert_delta}
\begin{figure}[htb]
\centering
\vspace{-1.5cm}
\epsfig{figure=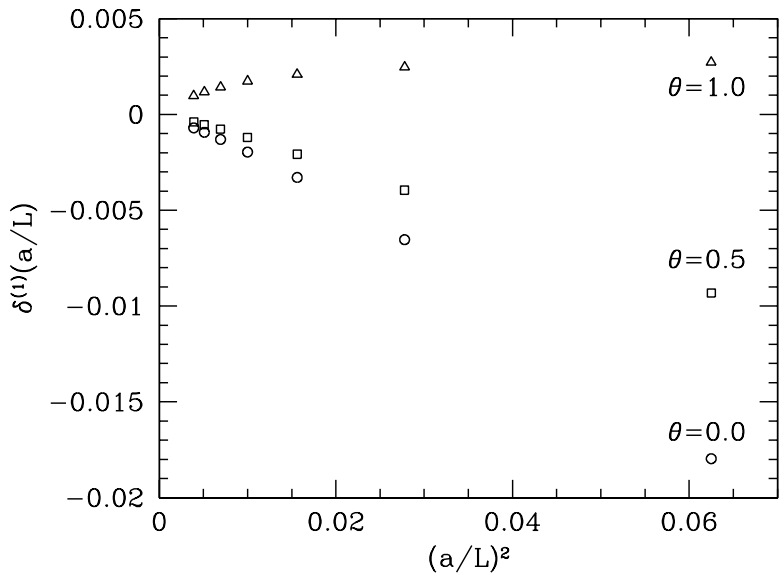,width=11.0cm}
\vspace{-1.75cm}
\caption{{\sl
Discretization errors of the step scaling function at one-loop level.
The continuum extrapolation of the non-perturbative results uses
$(a/L)^2 < 0.02$.
}}\label{f_delta}
\end{figure}

The one-loop expansion at hand is also helpful to study discretization 
errors in the step scaling function. 
Using \Eq{e_zastatexpansion}, the step
scaling function at lattice spacing $a$ is expanded as
\begin{equation}
\SigmaAstat(u,a/L)=1+u\,\SigmaAstatoneloop(a/L)+{\rm O}(u^2) \,,
\end{equation}
and its continuum limit $\sigmaAstat(u)$ as
\begin{equation}
\sigmaAstat(u)=
1+u\,\sigmaAstatoneloop + u^2\sigmaAstattwoloop + {\rm O}(u^3) \,,
\end{equation}
with 
\bea
 \sigmaAstatoneloop&=&\ln(2)\,\gamma_0 \,, \nonumber \\
 \sigmaAstattwoloop&=& \hlf\ln^2(2)\,\gamma_0^2
                       +\ln^2(2)\,b_0\gamma_0+\ln(2)\,\gamma_1 \,. 
 \label{e_s01}
\eea
As a measure for the discretization errors, we define
\begin{equation}
\deltaAstat(u,a/L)=
{{\SigmaAstat(u,a/L)-\sigmaAstat(u)}\over{\sigmaAstat(u)}}
\end{equation}
with a perturbative expansion
\begin{equation}
\deltaAstat(u,a/L)=\deltaAstatoneloop(a/L)\,u+{\rm O}(u^2) \,.
\end{equation}
The one-loop coefficient $\deltaAstatoneloop$ versus the lattice spacing
squared is shown in \Fig{f_delta} for different values of $\theta$.
%
%
In the range of lattice spacings where our non-perturbative calculation is 
performed, the discretization errors at one-loop level are smaller than 
$1\% \times u$, giving rise to the hope that also the 
non-perturbative discretization 
errors are reasonably small. A welcome feature of $\deltaAstatoneloop$ 
at $\theta=0.5$ is that it is entirely dominated by the leading 
$a^2/L^2$--term in the $a$--expansion.

%% file: apdxc.tex
\section{Continuum step scaling function and matching at $L=1.436\,r_0$}
\label{App_contSSF}
In this last appendix we briefly discuss our parametrization 
(i.e.~the interpolating fit) of the continuum step scaling function and 
some details on the calculation of $\zastat$ at the matching scale 
$2\Lmax=1.436\,r_0$.
\subsection{Fits and error determination in the scale evolution}
\label{App_contSSF_err}
As described in \App{App_latSSF_extrap}, the continuum step scaling 
function $\sigmaAstat(u)$ has been obtained by extrapolating the lattice
data on $\SigmaAstat(u,a/L)$ at fixed $u$ to the continuum limit.
The next step is now to solve the recursion specified through 
\Eqs{gbar_lmax} -- (\ref{recursion}). In practice this is done by first 
representing the results for $\sigmaAstat(u)$ in \Tab{tab:sigastat_res} 
by a fit and then solving the recursion, \Eq{recursion}, with
$\sigmaAstat(u)$ given by the fit function.

Guided by the analysis for the step scaling function of the pseudoscalar 
density $\sigmap$ in \Ref{msbar:pap1} and the perturbative expansion 
discussed in \App{App_pert_delta}, 
\be
\sigmaAstat(u)=
1+s_0u+s_1u^2+s_2u^3+\ldots+s_nu^{n+1}
\label{SSF_cont_fit}
\ee
is chosen as fit ansatz. 
The two non-trivial leading terms are restricted by perturbation theory,
\bea
s_0=\sigmaAstatoneloop \,, \quad
s_1=\sigmaAstattwoloop \,,
\eea
cf.~\Eqs{e_s01}.
Up to three additional free fit parameters were allowed for.
All of these fits represent the data in \tab{tab:sigastat_res} well, 
and we decided to quote the
two-parameter fit (the curve of which is shown in \Fig{fig:sigma_zastat1})
as the final result for the functional form of $\sigmaAstat$.
To check that the polynomial fits are stable, we also investigated 
fits where only $s_0$ or even no coefficient at all is constrained to its
perturbative value.
This leads to consistent results for $\sigmaAstat(u)$; particularly the
latter fit then reproduces the perturbative prediction for $s_0$.

Having chosen a definite expression for $\sigmaAstat(u)$, the solution of 
the associated recursion is unique. 
Since the errors on the step scaling function stem from different 
simulation runs and are hence uncorrelated, the errors on the fit 
parameters in the polynomial (\ref{SSF_cont_fit}) for $\sigmaAstat(u)$ and 
those on the $v_k$-s calculated from it can be estimated 
straightforwardly by the standard error propagation rules.
Finally, by increasing the number of free fit parameters (while fixing 
$s_0,s_1$ to perturbation theory) as mentioned above, we convinced 
ourselves that the systematic error induced by the choice of fit functions 
is well under control: in fact, we then observed the expected pattern of 
finding slightly different errors but compatible results at comparably good 
overall fit quality.
\subsection{Calculation of $\zastat$ at the low-energy matching scale}
\label{App_contSSF_match}
The total renormalization factor $\ZPhi$ introduced in \Sect{Sec_match}
involves the value of the renormalization constant $\zastat$ at our 
particular matching point: $\zastat(g_0,L/a)\big|_{L=1.436\,r_0}$.
As the latter connects a bare matrix element of the static-light axial
current to the one renormalized in the SF scheme, this amounts to calculate
$\zastat$ for a range of bare couplings commonly used in simulations in 
physically large volumes.

To extract $\zastat$ we exploit the fact that the required pairs
$(L/a,\beta)$ that match the condition $L/a=1.436\,r_0/a$ have already 
been determined for the relevant $\beta$--range in Appendix~C of 
\cite{msbar:pap1} by utilizing the known parametrization of $\ln(a/r_0)$ in 
terms of $\beta$ from \Ref{pot:r0_ALPHA}.
We thus could take over these pairs and computed $\zastat$ for $\theta=0.5$ 
from the renormalization condition (\ref{SFdef}) at the corresponding
values $\kappa=\hopc$ of the critical hopping 
parameter \cite{msbar:kapc_ct2l}.
The results for $\zastat(g_0,L/a)\big|_{L=1.436\,r_0}$ using the one-
and two-loop expressions for $\ct$, cf.~\Eq{ct_2loop}, are given 
in \Tab{tab:zastat1m_res}.
%
%
\begin{table}[htb]
\centering
\begin{tabular}{rccccccc}
\hline \\[-2.0ex]
&&& 
\multicolumn{2}{c}{$\ct^{{\rm 2-loop}}$} 
&& 
\multicolumn{2}{c}{$\ct^{{\rm 1-loop}}$} \\[1.0ex]
  $L/a$ & $\beta=6/g_0^2$ &&  
  $\kappa$ & $\zastat$ & & $\kappa$ & $\zastat$               \\[1.0ex]
\hline \\[-2.0ex]
  8  & 6.0219 && 0.13508 & 0.6926(15) && 0.13504 & 0.6932(10) \\
  10 & 6.1628 && 0.13565 & 0.6810(17) && 0.13564 & 0.6824(16) \\
  12 & 6.2885 && 0.13575 & 0.6786(18) && 0.13574 & 0.6795(16) \\
  16 & 6.4956 && 0.13559 & 0.6777(17) && 0.13558 & 0.6763(18) \\[1.0ex]
\hline \\[-2.0ex]
\end{tabular}
\caption{{\sl 
Results for $\zastat(g_0,L/a)$ at fixed scale $L=2\Lmax=1.436\,r_0$
with $\ct$ being set to either one- or two-loop.
The critical $\kappa$--values $\kappa=\hopc$ \protect\cite{msbar:kapc_ct2l} 
are the same as used for Table~C.1 of \protect\Ref{msbar:pap1}.  
}}\label{tab:zastat1m_res}
\end{table}
The difference originating from the two perturbative approximations for
$\ct$ is completely covered by the statistical errors so that we again 
consider $\ct^{{\rm 2-loop}}$ to already account for the dominant part
of the boundary cutoff effects in the gauge sector.
Similarly to the discussion in \App{App_latSSF_extrap}, the influence of 
the boundary improvement coefficient $\cttil$ in the fermionic sector can 
also be neglected at the level of our precision.
The parametrization of the results for 
$\zastat(g_0,L/a)\big|_{L=1.436\,r_0}$ by a polynomial fit in $(\beta-6)$, 
with $\ct^{{\rm 2-loop}}$ and $\castat$ from one-loop perturbation theory,
is quoted in \Sect{Sec_match}, where the coefficients in the first block
of \Tab{tab:polys} are to be combined with \Eq{e_ZApoly}.
The smooth dependence of $\zastat$ on $\beta$ in the studied region of bare
couplings suggests that this representation can also be slightly extended
down to $\beta=6.0$ (even though we could not directly simulate that point
for the same reason as in case of $\zp$ \cite{msbar:pap1}), and we
therefore regard it as a reliable representation of our data over the whole
range $6.0\le\beta\le 6.5$.

As in \App{App_latSSF_extrap}, we also set $\castat=0$ instead of one-loop
in the analysis of the data on $\zastat(g_0,L/a)\big|_{L=1.436\,r_0}$, and
the results are listed in the middle part of \Tab{tab:zastat1mcX0_res}.
%
%
\begin{table}[htb]
\centering
\begin{tabular}{rccccccc}
\hline \\[-2.0ex]
&&& 
\multicolumn{2}{c}{$\castat=0$} 
&& 
\multicolumn{2}{c}{$\csw=0$} \\[1.0ex]
  $L/a$ & $\beta=6/g_0^2$ &&  
  $\kappa$ & $\zastat$ & & $\kappa$ & $\zastat$               \\[1.0ex]
\hline \\[-2.0ex]
  4  & 5.6791 && ---     & ---        && 0.15268 & 0.6923(13) \\
  6  & 5.8636 && ---     & ---        && 0.15451 & 0.6315(16) \\
  8  & 6.0219 && 0.13508 & 0.6736(14) && 0.15341 & 0.6075(13) \\
  10 & 6.1628 && 0.13565 & 0.6633(17) && 0.15202 & 0.5964(18) \\
  12 & 6.2885 && 0.13575 & 0.6621(17) && 0.15078 & 0.5971(32) \\
  16 & 6.4956 && 0.13559 & 0.6627(17) && 0.14887 & 0.5991(52) \\[1.0ex]
\hline \\[-2.0ex]
\end{tabular}
\caption{{\sl 
Results for $\zastat(g_0,L/a)$ at fixed scale $L=2\Lmax=1.436\,r_0$
for $\castat=0$ and unimproved Wilson fermions (i.e.~$\csw=0$ and thus,
$\ca=\castat=0$ too), where $\ct$ was kept at its two-loop value.
}}\label{tab:zastat1mcX0_res}
\end{table}
In contrast to the step scaling function, which did not change appreciably
under this replacement of the value for $\castat$, we observe an effect of
about 3\% in $\zastat$ at the low-energy matching scale 
$L=2\Lmax=1.436\,r_0$.

Furthermore, we addressed the case of unimproved Wilson fermions by also
setting $\csw=0$ in the relativistic fermion action and, after having
computed the needed estimates of the critical hopping parameter for this
situation, carried out the additional runs to determine the renormalization 
constant.
In this case the pairs $(L/a,\beta)$ were extended to lower values of 
$\beta$ in order to be able to make contact with the $\beta$--region that
is typically employed in simulations to calculate the bare matrix element
defining the B-meson decay constant, 
as e.g.~those in \Refs{stat:fnal2,stat:DMcN94}.
The resulting estimates on $\zastat$ are shown in the right part 
of \Tab{tab:zastat1mcX0_res}, and the corresponding polynomial 
representations for both aforementioned cases are as well found 
via \Eq{e_ZApoly} together with the two lower blocks of \Tab{tab:polys}.

We conclude this discussion with the general remark that the uncertainties 
of the entering critical $\kappa$--values (of 1--2 and 2--4 on the last
decimal place in the case of $\csw=\mbox{non-perturbative}$ and $\csw=0$, 
respectively) do not affect the $Z$--factors significantly.